\documentclass[
a4paper,
aps,
prd,
twocolumn,
preprintnumbers,
superscriptaddress,
nofootinbib
]{revtex4-1}

\usepackage{fullpage}
\usepackage{amsfonts}
\usepackage{amsmath}
\usepackage{slashed}
\usepackage{amssymb}
\usepackage{graphicx}
\usepackage{epic}
\usepackage{eepic}
\usepackage{epsfig}
\usepackage{latexsym}
\usepackage{color}
\usepackage{float}
\usepackage{multirow}
\usepackage{hyperref}
\usepackage{relsize}
\usepackage{tabularx}
\usepackage[caption=false]{subfig}
\usepackage{diagbox}
\usepackage[left=1.90cm,right=1.90cm,top=1.90cm,bottom=1.90cm]{geometry}
\usepackage{shorthand}

\newcommand{\dmu}{\partial_\mu}
\newcommand{\dnu}{\partial_\nu}
\newcommand{\fmn}{F_{\mu\nu}}
\newcommand{\fmnup}{F^{\mu\nu}}
\newcommand{\hmn}{\widehat{F}_{\mu\nu}}
\newcommand{\hmnup}{\widehat{F}^{\mu\nu}}

\newcommand{\half}{\frac{1}{2}}

\begin{document}

\title{Varying gauge couplings and collider phenomenology}

\author{Ulf Danielsson}
\email{ulf.danielsson@physics.uu.se}
\affiliation{Department of Physics and Astronomy, Uppsala University, Box 516, SE-751 20 Uppsala, Sweden}

\author{Rikard Enberg}
\email{rikard.enberg@physics.uu.se}
\affiliation{Department of Physics and Astronomy, Uppsala University, Box 516, SE-751 20 Uppsala, Sweden}

\author{Gunnar Ingelman}
\email{gunnar.ingelman@physics.uu.se}
\affiliation{Department of Physics and Astronomy, Uppsala University, Box 516, SE-751 20 Uppsala, Sweden}
\affiliation{Swedish Collegium for Advanced Study, Thunbergsv\"agen 2, SE-752 38 Uppsala, Sweden}

\author{Tanumoy Mandal}
\email{tanumoy.mandal@physics.uu.se}
\affiliation{Department of Physics and Astronomy, Uppsala University, Box 516, SE-751 20 Uppsala, Sweden}

\date{May 27, 2019}

\begin{abstract}
In this paper we investigate a natural extension of the Standard Model that involves varying coupling constants. This is a general expectation in any fundamental theory such as string theory, and there are good reasons for why new physics could appear at reachable energy scales.  We investigate the collider phenomenology of models with varying gauge couplings where the variations are associated with real singlet scalar fields. We introduce three different heavy scalar fields that are responsible for the variations of the three gauge couplings of the Standard Model. This gives rise to many interesting collider signatures that we explore, resulting in exclusion limits based on the most recent LHC data, and predictions of the future discovery potential at the high-luminosity LHC.
\end{abstract}

\pacs{XXX}

\keywords{Varying couplings, Exclusion limits, Large Hadron Collider.}

\maketitle

\section{Introduction}
\label{sec:intro}

In a fundamental theory, there should be no free parameters. For example, in string theory, all coupling constants are given by vacuum expectation values (vevs) of scalar fields known as \emph{moduli fields}. In top-down constructions of low-energy effective theories from string theory, the fundamental constants should in principle be derived from such vevs at a high scale, but there are many obstacles involved in constructing the Standard Model (SM), or extensions thereof, in this way. In this paper, we instead want to consider the idea of couplings set by scalar fields in bottom-up constructions, where the low-energy theory is constructed at lower scales as an effective field theory with a cutoff scale. We will here consider the Standard Model, where all  dimensionless parameters could in principle be controlled by scalar fields.

One can make the case that this argument in favor of varying coupling constants is as strong as the theoretical arguments in favor of supersymmetry. Just as with supersymmetry, it is based on a very simple and reasonable assumption about the properties of the underlying fundamental theory. Estimating the relevant energy scales for these extensions is much more difficult. In the case of supersymmetry, it has been argued through naturalness that supersymmetry at low energies could solve the observed hierarchy of scales. In the case of varying couplings, the observed existence of a low energy scalar such as the Higgs boson can be taken as an argument suggesting that the scalars we study in this paper could have reasonably low masses as well. The Higgs boson is special in the sense that it controls the breaking of a gauge symmetry, but in addition it controls the masses of the standard model particles. Normalizing with the help of the Planck mass---which is an obvious thing to do in a fundamental theory---it means that the Higgs boson is better thought of as controlling a set of dimensionless parameters in the Standard model. This is similar to the scalars we have in mind. We cannot see any strong reason for why these other scalars must have masses way beyond the Higgs mass. Our conclusion is that there are good arguments in favor of conducting searches for such particles at the LHC and beyond.

There have been many studies of models where the coupling constants change over time~\cite{Uzan:2010pm}. In particular, models with a varying $\alpha_\text{EM}$ have been widely discussed; in these models the associated scalar field is light and $\alpha_\text{EM}$ has not been constant over the history of the Universe. Today the constraints on this kind of coupling variation of $\alpha_\text{EM}$ are very strong, suggesting the mass of any associated scalar field to be very high.  This is not the kind of variation we have in mind in this paper. Here we are rather interested in the particle physics implications of dynamical couplings in terms of new scalar particles associated with the scalar fields.

Even if the effects of varying couplings can only be seen at high energies, they can still be of crucial importance for cosmology. This includes varying Yukawa couplings leading to a strong electroweak phase transition \cite{Baldes:2016rqn,Braconi:2018gxo} or recently a varying strong coupling $\alpha_S$ leading to QCD confinement in the early Universe and possible scenarios for baryogenesis \cite{Ipek:2018lhm}. The latter work introduced a nontrivial potential for the scalar field governing the strength of the strong interaction. In this way, it was possible in \cite{Ipek:2018lhm} to obtain a temperature-dependent vev for the scalar field. This, in turn, allowed for QCD to remain confined at very high temperatures, which might help explain baryogenesis. In [3], it was natural to pick the energy scales of the new physics to be of order TeV. Our general analysis outlines how data from LHC may be used to put meaningful constraints on such models.

An early, consistent varying coupling model was proposed by Jacob Bekenstein in 1982 \cite{Bekenstein:1982eu,Bekenstein:2002wz}. This was a model for electromagnetism where the electromagnetic coupling $e$ is given by a light scalar field. The model is gauge invariant and invariant under rescaling of the scalar field, and reduces to ordinary electromagnetism if the scalar field reduces to a constant value. We will refer to this model as the Bekenstein model (BM) and will discuss it further in 
Sec.~\ref{sec:model}.

The Bekenstein model was proposed in the context of cosmology and has so far only been considered with this purpose. Recently, however, we introduced the idea~\cite{Danielsson:2016nyy} that the scalar field in the Bekenstein model could be heavy, with a mass on the TeV scale, such that it could potentially be discovered at the LHC. In~\cite{Danielsson:2016nyy}, we concentrated on the electromagnetic coupling and the corresponding scalar field that couples to photons. 

In this paper, we generalize this model. In the SM, the three gauge couplings $g_1,g_2,g_3$ of the gauge group  
$G_{\mathrm{SM}}=\mathrm{U}(1)_Y\times\mathrm{SU}(2)_L\times\mathrm{SU}(3)_C$ as well as the Yukawa couplings of the fermions, the CKM matrix, and the Higgs quartic self-coupling $\lambda_H$ should be controlled by scalar fields. We will here as a first step restrict ourselves to the electroweak theory and will write down the model where the $\mathrm{U}(1)_Y\times\mathrm{SU}(2)_L$ gauge couplings $g_1$ and $g_2$ and the strong coupling $g_3$ are set by scalar fields. We will briefly discuss the further generalization to the Higgs and Yukawa sector, but will save the detailed implementation for future work. Ultimately, one would like to replace all dimensionless constants of the SM with values set by their corresponding scalar fields.

The paper is organized as follows. In Sec.~\ref{sec:model}, we briefly discuss the construction of
varying fine-structure constant theory. Following the same method, we discuss in detail the varying electroweak gauge
couplings and varying strong coupling models. In these models, the variations of three gauge couplings of the SM are promoted to three different heavy scalar fields. In Sec.~\ref{sec:pheno}, we show the branching ratios (BRs) and production cross 
sections of these scalars at the LHC. We derive exclusion limits of the free parameters of these models in
Sec.~\ref{sec:exclu} by using the latest LHC data. In Sec.~\ref{sec:reach}, we estimate the projected reach at the
high-luminosity LHC (HL-LHC). Finally, we summarize and conclude our results in Sec.~\ref{sec:conclu}.

\section{The model}
\label{sec:model}

In Ref.~\cite{Danielsson:2016nyy}, we considered the model of varying 
electromagnetic (EM) coupling as our starting point. In this model, the  
EM coupling is not a fundamental constant but a function of space-time, i.e., 
$e(x) = e\ve(x)$, where $\ve(x)$ is a dimensionless scalar field and where we have, for later convenience, extracted the vev 
$\langle e(x)\rangle=e$ such that $\ve=1$ at the minimum of the scalar potential.
We use the following notation for the rest of the paper: a space-time varying constant $\kp(x)$ will be denoted by 
$\widetilde{\kp}\equiv \kp(x)$ and its vev by $\langle \widetilde{\kp}\rangle=\kp$. For instance, in this notation $\widetilde{e}\equiv e(x)$ and $e$ is the constant value that we measure in vacuum. We also omit the argument ``$x$'' from a field for ease of notation.
The field $\ve$ has noncanonical kinetic term of the form
\begin{equation}
\frac{1}{2}\frac{\Lambda^2}{\ve^2} (\dmu\ve)^2,
\end{equation}
where $\Lambda$ is an energy scale. This form of the kinetic term respects rescaling invariance, and is typical of kinetic terms for moduli fields in string theory.

For particle physics applications it is convenient to deal with a scalar field $\phi$ that has physical mass dimension one and a canonical kinetic term $(1/2)(\dmu\phi)^2$, instead of the noncanonically normalized $\ve$. This can be done by first 
writing $\ve=\exp\lt(\varphi\rt)$, with $\varphi=\ln\lt(\widetilde{e}/e\rt)$, and then rescaling by an energy scale $\Lambda$ such that $\varphi = \phi/\Lambda$. The field $\ve$ is expanded to the first order as $\ve \simeq 1+\varphi = 1+\phi/\Lambda$.

Bekenstein showed in \cite{Bekenstein:1982eu} that the theory defined using the field strength tensor
\begin{equation}
\hmn = \frac{1}{\ve} \left[ \dmu (\ve A_\nu) - \dnu (\ve A_\mu) \right],
\end{equation}
accompanied by a modified gauge transformation of the field $A_\mu$ given by
\begin{equation}
 \ve A_\mu \to \ve A_\mu + \dmu\alpha
\label{eq:gaugetransf}
\end{equation}
is gauge and rescaling invariant. In particular, this means that in the ordinary QED theory, $\widetilde{e}A_\mu$ is always replaced by 
$e\ve A_\mu$. 
We define the ordinary QED field strength tensor as $\fmn=\dmu A_\nu - \dnu A_\mu$.

In \cite{Danielsson:2016nyy}, we derived the form of the modified QED Lagrangian suitable for particle physics analyses with the varying coupling $\widetilde{e}$ replaced by the constant coupling $e$ and with couplings to the real scalar field $\phi$ of mass $m_\phi$ introduced. To illustrate the main construction, let us consider a simple $\mathrm{U}(1)_{\textrm{EM}}$ theory
consisting of photon and a fermion $f$ with EM charge $eQ_f$. This derivation is briefly described in Appendix \ref{app:derivation}. The resulting new terms for interactions of the photon with the fermion are (there are also interactions of the scalar with $W^\pm$, but we do not list these terms here; see~\cite{Danielsson:2016nyy})
\begin{align}
\label{eq:QEDlag}
{\cal L} &\supset \half(\dmu\phi)^2 - \half m_\phi^2\phi^2 - \frac{1}{\Lambda} \dmu\phi \, A_\nu  \fmnup  \nn\\ 
&+ \frac{e Q_f}{\Lambda} \phi\, \bar{f} \gamma^\mu f \, A_\mu + \mathcal{O}\lt(\frac{1}{\Lm^2}\rt).
\end{align}
Notice that the kinetic term for the photon field and the interaction term of the scalar with photons are not individually gauge invariant, but their sum is. We may use integration by parts and the equations of motion for $\fmn$ to remove the last term and rewrite the next-to-last term so that there is no direct coupling of the type $\phi\bar{f}f\gamma$. The Lagrangian then instead takes the more familiar form (see Appendix \ref{app:alternative} for details)
\begin{align}
\label{eq:QEDlag2}
{\cal L} &\supset \half(\dmu\phi)^2 - \half m_\phi^2\phi^2  \nn\\
&+ \frac{1}{2\Lambda}\phi\fmn\fmnup + \mathcal{O}\lt(\frac{1}{\Lm^2}\rt), 
\end{align}
with the well-known dimension-five operator coupling a singlet scalar to a gauge field. This form of the Lagrangian is related to (\ref{eq:QEDlag}) by a field redefinition, and the two forms are equivalent  in the sense that they predict the same $S$-matrix elements, although the explicit amplitudes will appear different.

The $\phi\fmn\fmnup$ term in Eq.~(\ref{eq:QEDlag2}) gives the new interactions in the model. The obtained theory is therefore regular QED, plus a real scalar field that couples to photons through a dimension-five operator, where interestingly the coupling is given by the inverse of the scale $\Lambda$ and no other parameters.


\subsection{Varying electroweak couplings}
\label{sec:EWmodel}

Above we discussed the model with varying EM coupling $e$. Now, we generalize this 
picture to varying gauge couplings in the electroweak (EW) sector. This 
generalization has been previously studied in Refs.~\cite{Kimberly:2003rz,Shaw:2004hk} in
the context of cosmology. Here, our primary aim is to study the phenomenology of this model 
in the context of the LHC. We let the two gauge couplings $g_1$ and $g_2$ of the EW
sector (where $g_1$ and $g_2$ are the $\mathrm{U}(1)_Y$ and $\mathrm{SU}(2)_L$ gauge couplings, 
respectively) be specified by scalar fields. We consider two cases: 
one scalar theory (1ST), where both couplings are set by the same scalar field, and
two scalar theory (2ST), where each coupling is set by a different field.

In 1ST, a single scalar field $S_0$ is responsible for the variation of both the 
couplings $g_1$ and $g_2$. It is more natural to introduce two different scalar fields 
responsible for the variations of the two independent gauge couplings---this is the 2ST model which we are
mainly interested in. However, for pedagogical reasons, we first discuss the 1ST model, due to its simple nature 
with only two free parameters.

\subsubsection{One scalar theory}

In the 1ST, the space-time varying gauge couplings are given by $\widetilde{g}_i=g_i\,\ve$, 
for $i=1,2$, where $\ve$ is the new scalar field that we write as 
$\ve=\exp\lt(S_0/\Lambda_0\rt)\simeq 1+S_0/\Lambda_0$
by introducing a canonically normalized scalar field $S_0$ and the new physics scale $\Lm_0$. This is an obvious generalization of the 
formalism described before for varying EM coupling. 
Similar to Eq.~\eqref{eq:QEDlag2}, the interaction Lagrangian before EWSB can be written as 
\begin{align}
\label{eq:efflagEW1} 
\mc{L} \supset  \frac{1}{2\Lm_0}S_0\lt(\mc{B}^2 + \mc{W}^{2}\rt),
\end{align}
where $\mc{B}^2=B_{\mu\nu}B^{\mu\nu}$ and $\mc{W}^{k2}=W_{\mu\nu}^kW^{k\mu\nu}$ (with $k=1,2,3$) and
$B_{\mu\nu}$ and $W^k_{\mu\nu}$ are the field-strength tensors for 
the $\mathrm{U(1)}_Y$ and $\mathrm{SU(2)}_L$ 
gauge bosons, respectively. These interactions are nonrenormalizable effective dimension-five operators suppressed by the 
scale $\Lm_0$.
The scalar potential of this theory consists of the SM Higgs doublet $\Phi$ and the new scalar $S_0$, and the most
general form is given by 
\begin{align}
V(\Phi,S_0) &= \mu^2\lt(\Phi^\dagger\Phi\rt) + \lm\lt(\Phi^\dagger\Phi\rt)^2\nn\\
&+ \sum_{n=1}^2 \lm_n S_0^n \lt(\Phi^\dagger\Phi\rt) + \sum_{n=1}^4 \lm^{\prime}_n S_0^n,
\end{align}
where $\mu,\lm,\lm_n$, and $\lm^{\prime}_n$ are the free couplings of the potential. In this
theory, $S_0$ will not get a vev\footnote{In general, for an arbitrary potential, it is possible to have the global minimum 
away from $S_0=0$. However, one can always remove the vev by redefining $S_0$.}, but the Higgs doublet $\Phi$ gets a vev and breaks the EW symmetry as usual in the SM. 
After EWSB, an off-diagonal term $S_0h$ ($h$ is the Higgs field) in the scalar mass matrix is generated from the $\lm^{\prime}_1 S\lt(\Phi^\dagger\Phi\rt)$
interaction. This leads to the mixing between the Higgs $h$ and the new scalar $S_0$. This mixing is
severely constrained from LHC measurements~\cite{Khachatryan:2016vau,Sirunyan:2018koj,ATLAS:2019slw}.
Therefore, for simplicity, we do not consider the mixing of the new scalar with the Higgs and assume
$S_0$ is the physical mass eigenstate.
Note that the Weinberg angle $\theta_w=\arctan(\widetilde{g}_1/\widetilde{g}_2)$ is not varying, since the variations of $\widetilde{g}_1$ and $\widetilde{g}_2$ are the same and 
cancel in the ratio. However, the gauge boson masses are varying, with variations given by
\begin{align}
\label{eq:mdyn1}
\widetilde{m}_Z^2 &= \frac{1}{4}v^2\left(\widetilde{g}_1^2+\widetilde{g}_2^2\right)= m_{Z}^2 \left(1 + \frac{2S_0}{\Lambda_0} \right)
+ \mathcal{O}\lt(\frac{1}{\Lm_0^2}\rt)\nn\\
\widetilde{m}_W^2 &= \frac{1}{4}v^2\widetilde{g}_2^2 
= m_{W}^2 \left(1 + \frac{2S_0}{\Lambda_0}\right) + \mathcal{O}\lt(\frac{1}{\Lm_0^2}\rt).
\end{align}
These expressions for the gauge boson masses must be inserted in the gauge boson-Higgs Lagrangian coming from the covariant derivative in the kinetic term for the Higgs field,
\be
\label{eq:efflagH}
\mc{L} \supset \left[\frac{1}{2} \widetilde{m}_Z^2 Z^\mu Z_\mu + \widetilde{m}_W^2 W^{\mu+} W_\mu^-  \right] \left(1+\frac{h}{v}\right)^2,
\ee
which leads to vertices of the type $S_0VV$, $S_0hVV$, and $S_0hhVV$.

To rewrite the interactions after EWSB, we insert $B_{\mu} = c_wA_{\mu}-s_wZ_{\mu}$ 
and $W^3_{\mu} = s_wA_{\mu}+c_wZ_{\mu}$ (where $s_w=\sin\theta_w$ and $c_w=\cos\theta_w$) in Eq.~\eqref{eq:efflagEW1}, and 
replace $\widetilde{m}_Z$ and $\widetilde{m}_W$ in Eq.~\eqref{eq:efflagH} by the corresponding expressions
from Eq.~\eqref{eq:mdyn1}. After this replacement, we obtain the following interactions of $S_0$ with the physical gauge bosons:
\begin{align}
\mc{L} &\supset  \frac{1}{2\Lm_0}S_0\left[\mc{F}^2 
+\lt(\mc{Z}^2+ 2m_{Z}^2Z_\mu Z^\mu\rt) \right. \nn\\ 
&+ \left. 2\lt(\mc{W}^{+}\mc{W}^{-} + 2m_{W}^2 W_{\mu}^{+} W^{-\mu}\rt)\right]+\dots ,
\end{align}
where $\mc{V}\equiv\mc{V}_{\mu\nu}= \partial_\mu \mc{V}_\nu - \partial_\nu \mc{V}_\mu$ is the Abelian field-strength tensor for the gauge boson, $\mc{V} = \lt\{\mc{F},\mc{Z},\mc{W}^{\pm}\rt\}$.
The \emph{dots} stand for terms with four- and five-point interactions that arise from the 
non-Abelian part of the $W^k_{\mu\nu}$ field-strength tensor of Eq.~\eqref{eq:efflagEW1}.
We do not explicitly show these interactions here, but they are included and are important to preserve the gauge
invariance of the theory. The interactions of $S_0$ with the Higgs associated with two vector bosons
can be readily obtained from Eq.~\eqref{eq:efflagH}.

\subsubsection{Two scalar theory}

In 2ST, the space-time variations of the two gauge couplings in the EW sector are given by two new scalars, 
$\widetilde{g}_i=g_{i}\,\ve_i$, for $i=1,2$, where 
$\ve_i=\exp\lt(S_i/\Lambda_i\rt)\simeq 1+S_i/\Lambda_i$.
Here, we introduce two new, canonically normalized fields $S_i$ and two corresponding scales 
$\Lambda_i$. As before, the scalars $S_i$ will not get vevs,
and mixing between $S_i$ and the Higgs can arise from the general scalar potential. Therefore, the
$S_i$ fields are not, in general, physical mass eigenstates, but interaction eigenstates. 
Although we neglect the mixing of $S_i$ with the Higgs to satisfy the Higgs measurements, we do consider
mixing between the two new scalars as there is no \emph{a priori} reason to neglect 
$S_1\leftrightarrow S_2$ mixing. This mixing can lead to interesting collider signatures which we will see later.

There will be two sources of interactions of $S_i$ with the SM fields. The first source is 
the interaction terms similar to Eq.~(\ref{eq:efflagEW1}) as follows: 
\begin{align}
\label{eq:efflagEW2} 
\mc{L} \supset  \frac{1}{2\Lm_1}S_1 \mc{B}^2 + \frac{1}{2\Lm_2}S_2 \mc{W}^2.
\end{align}

\begin{table*}
\centering
\caption{\label{tab:coup} Couplings of $\phi_i$ to two gauge bosons in the 2ST model. The corresponding Lagrangian
terms are shown in Eq.~\eqref{eq:effint}.}
\begin{tabularx}{\textwidth}{@{}l *3{>{\centering\arraybackslash}X} r@{}}
\hline\hline
For & $\kp_{\gm\gm}^i$ & $\kp_{\gm Z}^i$ & $\kp_{ZZ}^i$ & $\kp_{WW}^i$ \\ 
\hline 
$\phi_1$ & $\displaystyle +\frac{1}{2}\lt(\frac{c_{\al}c_w^2}{\Lm_1}+\frac{s_{\al}s_w^2}{\Lm_2}\rt)$ & $\displaystyle - s_wc_w\lt(\frac{c_{\al}}{\Lm_1}-\frac{s_{\al}}{\Lm_2}\rt)$ & $\displaystyle +\frac{1}{2}\lt(\frac{c_{\al}s_w^2}{\Lm_1}+\frac{s_{\al}c_w^2}{\Lm_2}\rt)$ & $\displaystyle +\frac{s_{\al}}{\Lm_2}$ \\ \\
$\phi_2$ & $\displaystyle -\frac{1}{2}\lt(\frac{s_{\al}c_w^2}{\Lm_1}-\frac{c_{\al}s_w^2}{\Lm_2}\rt)$ & $\displaystyle +s_wc_w\lt(\frac{s_{\al}}{\Lm_1}+\frac{c_{\al}}{\Lm_2}\rt)$ & $\displaystyle -\frac{1}{2}\lt(\frac{s_{\al}s_w^2}{\Lm_1}-\frac{c_{\al}c_w^2}{\Lm_2}\rt)$ & $\displaystyle +\frac{c_{\al}}{\Lm_2}$ \\ 
\hline\hline
\end{tabularx} 
\end{table*}

The second source of interactions comes from the dynamical nature of the Weinberg angle and the gauge boson masses. 
The tangent of the Weinberg angle $\widetilde{\theta}_w$ (denoted by $\widetilde{t}_w$) is given by
\begin{align}
\widetilde{t}_w = \frac{\widetilde{g}_1}{\widetilde{g}_2}
=t_{w} \exp\lt(\frac{S_1}{\Lambda_1} - \frac{S_2}{\Lambda_2}\rt),
\end{align}
such that the sine ($\widetilde{s}_w$) and cosine ($\widetilde{c}_w$) are given by
\begin{align}
\widetilde{s}_w &= s_{w} \left[ 1 + c_{w}^2 \left(\frac{S_1}{\Lambda_1}-\frac{S_2}{\Lambda_2}\right)\right] 
+ \mathcal{O}\lt(\frac{1}{\Lm_i^2}\rt)\nn\\
\widetilde{c}_w &= c_{w} \left[ 1 - s_{w}^2 \left(\frac{S_1}{\Lambda_1}-\frac{S_2}{\Lambda_2}\right)\right] + \mathcal{O}\lt(\frac{1}{\Lm_i^2}\rt).
\label{eq:weindyn}
\end{align}
The squares of the gauge boson masses are then expressed as
\begin{align}
\label{eq:mdyn2}
\widetilde{m}_Z^2 &= \frac{1}{4}v^2\left(\widetilde{g}_1^2+\widetilde{g}_2^2\right) \nn\\
&= m_{Z}^2 \left[ 1 + 2 \left(s_{w}^2 \frac{S_1}{\Lambda_1} + c_{w}^2\frac{S_2}{\Lambda_2} \right) \right]
+ \mathcal{O}\lt(\frac{1}{\Lm_i^2}\rt)\nn\\
\widetilde{m}_W^2 &= \frac{1}{4}v^2\widetilde{g}_2^2 
= m_{W}^2 \left( 1 + \frac{2S_2}{\Lambda_2}\right) + \mathcal{O}\lt(\frac{1}{\Lm_i^2}\rt).
\end{align}
It is interesting to note that in this model the Fermi coupling $G_F$ is \textit{not} dynamical, 
\begin{equation}
 \frac{G_F}{\sqrt{2}}  = \frac{\widetilde{g}_2^2}{8\widetilde{m}_W^2} = \frac{g_{2}^2}{8m_{W}^2}.
\end{equation}
This is because the parameters in the Higgs potential do not vary in this model. The mass parameter $\mu^2$ is dimensionful and is not allowed to vary, and since in this paper we focus on the gauge couplings we have not yet implemented a varying Higgs coupling $\lambda_H$, but this should be considered in the future and would lead to interesting phenomenology.

\begin{figure}[!h]
\includegraphics[height=3.2cm,width=4.2cm]{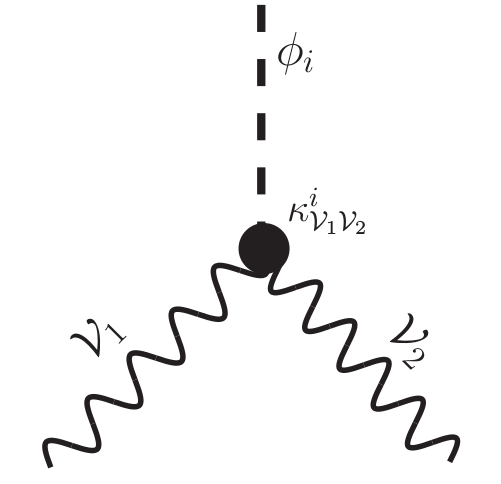}
\caption{A representative Feynman diagram for the $\phi_i\mc{V}_1\mc{V}_2$ interaction. 
The couplings $\kp_{\mc{V}_1\mc{V}_2}^i$
for this vertex are shown in Table~\ref{tab:coup}.}
\label{fig:phiVV}
\end{figure}

To rewrite the interactions after EWSB, we replace, in Eq.~\eqref{eq:efflagEW2}, $B_{\mu} = \widetilde{c}_wA_{\mu}-\widetilde{s}_wZ_{\mu}$ 
and $W^3_{\mu} = \widetilde{s}_wA_{\mu}+\widetilde{c}_wZ_{\mu}$. 
The expressions in Eq.~\eqref{eq:weindyn} must also be inserted in the above linear combinations.
This yields vertices with one or two new scalars and multiple gauge bosons. The vertices with more than one scalar are suppressed by higher orders of $1/\Lambda_i$, and we neglect these terms in the rest of the paper.
The fields $S_i$ are, in general, not mass eigenstates, which could be linear 
combinations of them.
We, therefore, define the mass eigenstates $\phi_i$ by the following linear combinations:
\begin{equation}
\begin{pmatrix} 
S_1 \\
S_2 
\end{pmatrix} =
\begin{pmatrix} 
\cos\al & -\sin\al \\
\sin\al & \cos\al 
\end{pmatrix}
\begin{pmatrix} 
\phi_1 \\
\phi_2 
\end{pmatrix}
\label{eq:licomb}
\end{equation}
where $\al$ is a mixing angle determined from the free
parameters of the scalar potential. 
As before, we replace $\widetilde{m}_Z$ and $\widetilde{m}_W$ in Eq.~\eqref{eq:efflagH} by the corresponding expressions from Eq.~\eqref{eq:mdyn2}. 
Finally, we collect the leading interactions of $\phi_i$ with two gauge bosons in the 
following Lagrangian:
\begin{align}
\label{eq:effint}
\mc{L} &\supset \phi_i\left[ \kp_{\gm\gm}^i\mc{F}^2 + \kp_{\gm Z}^i\mc{F}\mc{Z}
+\kp_{ZZ}^i\lt(\mc{Z}^2 + 2m_{Z}^2Z_\mu Z^\mu\rt) \nn \right. \\
&+ \left. \kp_{WW}^i\lt(\mc{W}^{+}\mc{W}^{-} + 2m_{W}^2 W_{\mu}^{+} W^{-\mu}\rt)\right]+\dots ,
\end{align}
where $\kp_{\mc{V}_1\mc{V}_2}^i$ represents the dimensionful couplings of $\phi_i$ with 
$\mc{V}_1\mc{V}_2=\lt\{\gm\gm,\gm Z,ZZ,WW\rt\}$ gauge boson pairs 
given in Table~\ref{tab:coup}. A representative Feynman diagram for the 
$\phi_i\mc{V}_1\mc{V}_2$ interaction originating from Eq.~\eqref{eq:effint} is shown in Fig.~\ref{fig:phiVV}.

\subsection{Varying strong coupling}
\label{subsec:QCDmodel}

Moving on to the strong interaction, we promote the strong coupling $g_3$ to a field the same way as done for the other interactions above, so that $\widetilde{g}_3 = g_{3}\,\ve_3$, and we have a new scalar $S_3$ defined through 
$\ve_3=\exp(S_3/\Lm_3)\simeq 1+S_3/\Lm_3$.
The interaction of the scalar with gluons is described by the dimension-five interaction
\begin{align}
\label{eq:efflagQCD} 
\mc{L} \supset  \frac{1}{2\Lm_3}S_3 G^a_{\mu\nu}G^{a\mu\nu}\ ,
\end{align}
where $G^a_{\mu\nu}$ with $a=1,\dots,8$ is the usual gluon field strength tensor. 
This new interaction gives rise to
$S_3 gg$, $S_3 g^3$, and $S_3 g^4$ vertices at order $\Lambda_3^{-1}$.
The scalar will interact with quarks in the same way as the Bekenstein scalar in the BM model interacts with electrons, as described in Sec.~\ref{sec:model}, i.e., in the original formulation there will be a four-point vertex $S_3 \bar q q g$, but after the field redefinition is performed, there will not be any direct coupling to fermions. Instead, it only couples via internal particles in amplitudes.
In general, $S_3$ can also mix with other scalars in the theory through the interactions in the
general scalar potential and lead to many interesting signatures at the colliders. However, a systematic study will be extremely complicated due to large number of free parameters in that situation.
Therefore, for simplicity, we discuss the varying strong coupling (VSC) theory with no mixing with the $S_1$ and $S_2$ scalars in the EW sector.


\section{Branching ratios and cross sections}
\label{sec:pheno}

For the phenomenological analysis, we implement the dimension-five effective interactions 
shown in Eqs.~\eqref{eq:efflagH}, \eqref{eq:efflagEW2}, and \eqref{eq:efflagQCD} in addition to the 
SM Lagrangian in \textsc{FeynRules}~\cite{Alloul:2013bka}, to obtain the 	
Universal FeynRules Output~\cite{Degrande:2011ua} model files suitable for the 
{\sc MadGraph}~\cite{Alwall:2014hca} event generator. The fields $S_1$ and $S_2$
are not the physical ones, and they are replaced by the linear combinations of
the physical states $\phi_1$ and $\phi_2$ as given in Eq.~\eqref{eq:licomb}
to obtain the interactions in the mass basis. All possible new interactions that
originate from these replacements are implemented in \textsc{FeynRules}, although
not all of them are shown in Eq.~\eqref{eq:effint}.
For photon initiated processes, we use the photon parton distribution function (PDF)
obtained using the approach in~\cite{Harland-Lang:2016apc,Harland-Lang:2016qjy} while
for other partons, the MMHT14LO~\cite{Harland-Lang:2014zoa} PDF set is used.
We use the default {\sc MadGraph} dynamical factorization and renormalization scales in our analysis. 
Events are showered and hadronized using {\sc Pythia8}~\cite{Sjostrand:2007gs}. 
Detector environment is simulated using {\sc Delphes}~\cite{deFavereau:2013fsa}, which uses the {\sc FastJet}~\cite{Cacciari:2011ma} package for jet clustering. 
The anti-$k_T$ algorithm~\cite{Cacciari:2008gp} is used for jet clustering with radius parameter $R=0.4$. 

\subsection{Branching ratios}
\label{ssec:BR}

\begin{figure}
\includegraphics[height=6.0cm,width=7.5cm]{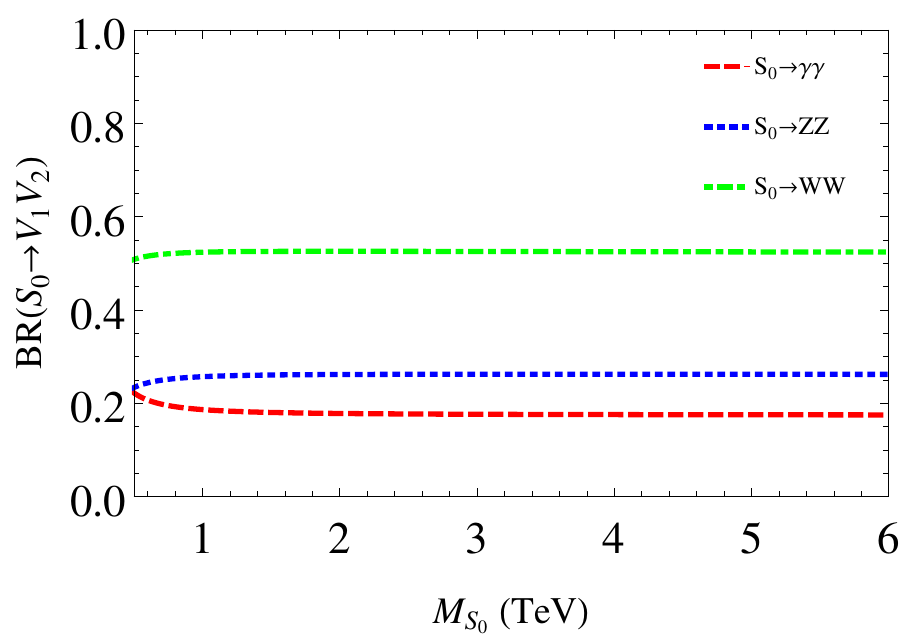}
\caption{Branching ratios of $S_0$ in the one scalar theory, as functions of $M_{S_0}$.}
\label{fig:BRphi}
\end{figure}
\begin{figure*}
\subfloat[]{\includegraphics[height=3.5cm,width=3.8cm]{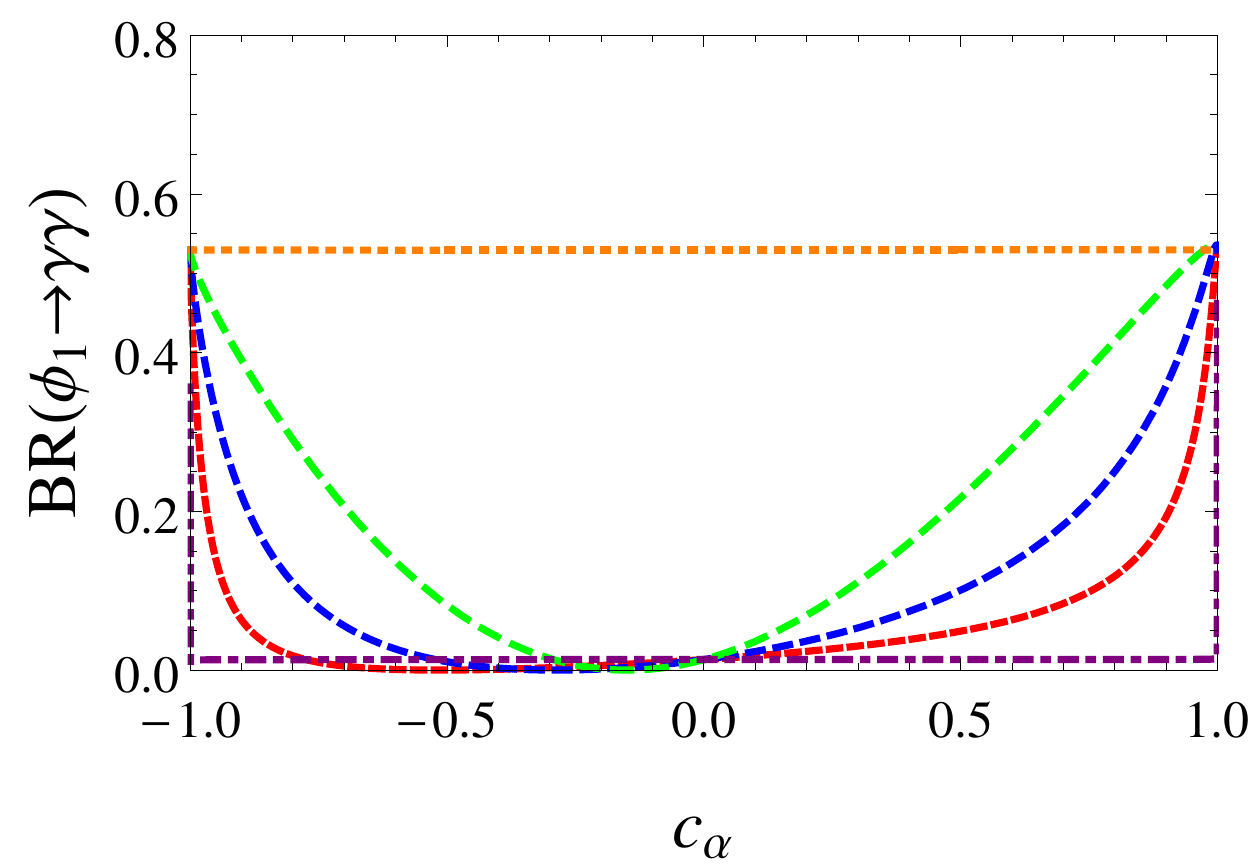}}\hspace{0.1cm}
\subfloat[]{\includegraphics[height=3.5cm,width=3.8cm]{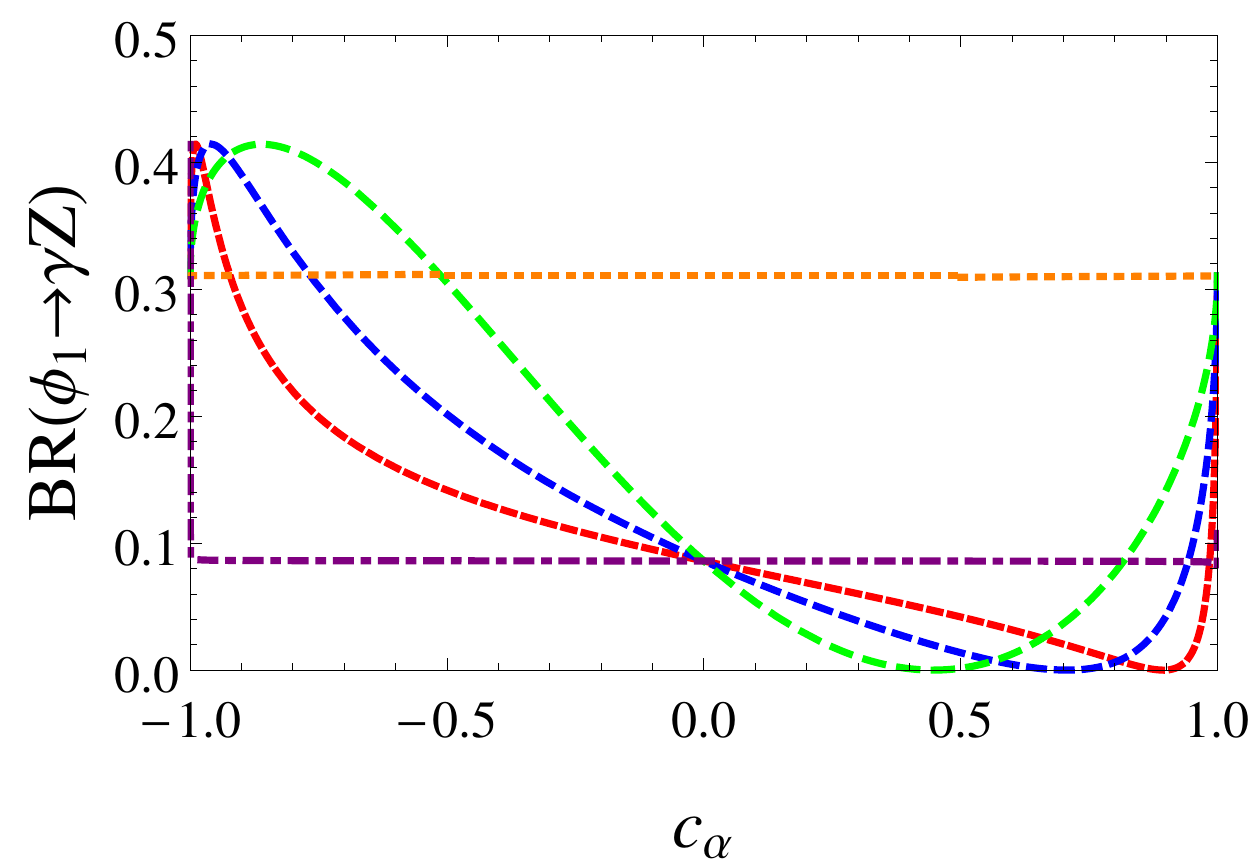}}\hspace{0.1cm}
\subfloat[]{\includegraphics[height=3.5cm,width=3.8cm]{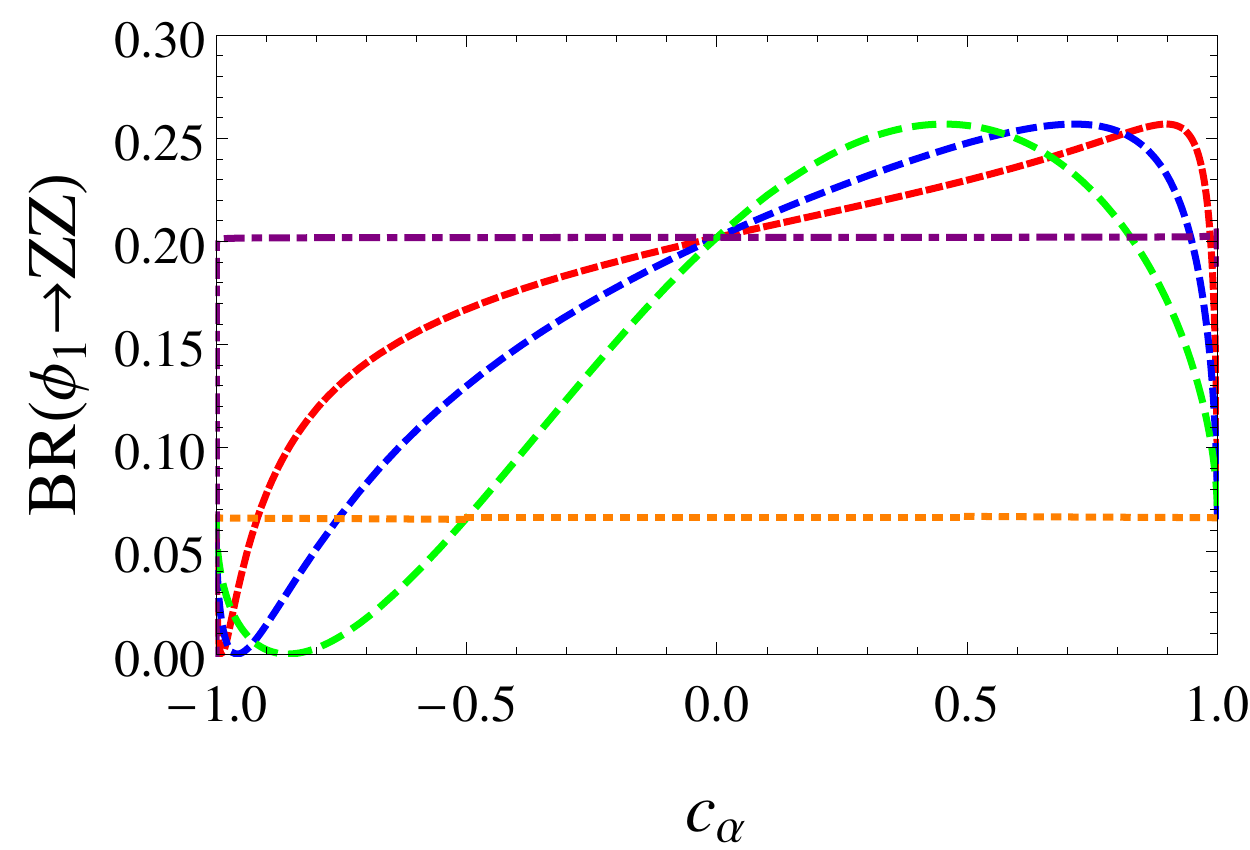}}\hspace{0.1cm}
\subfloat[]{\includegraphics[height=3.5cm,width=3.8cm]{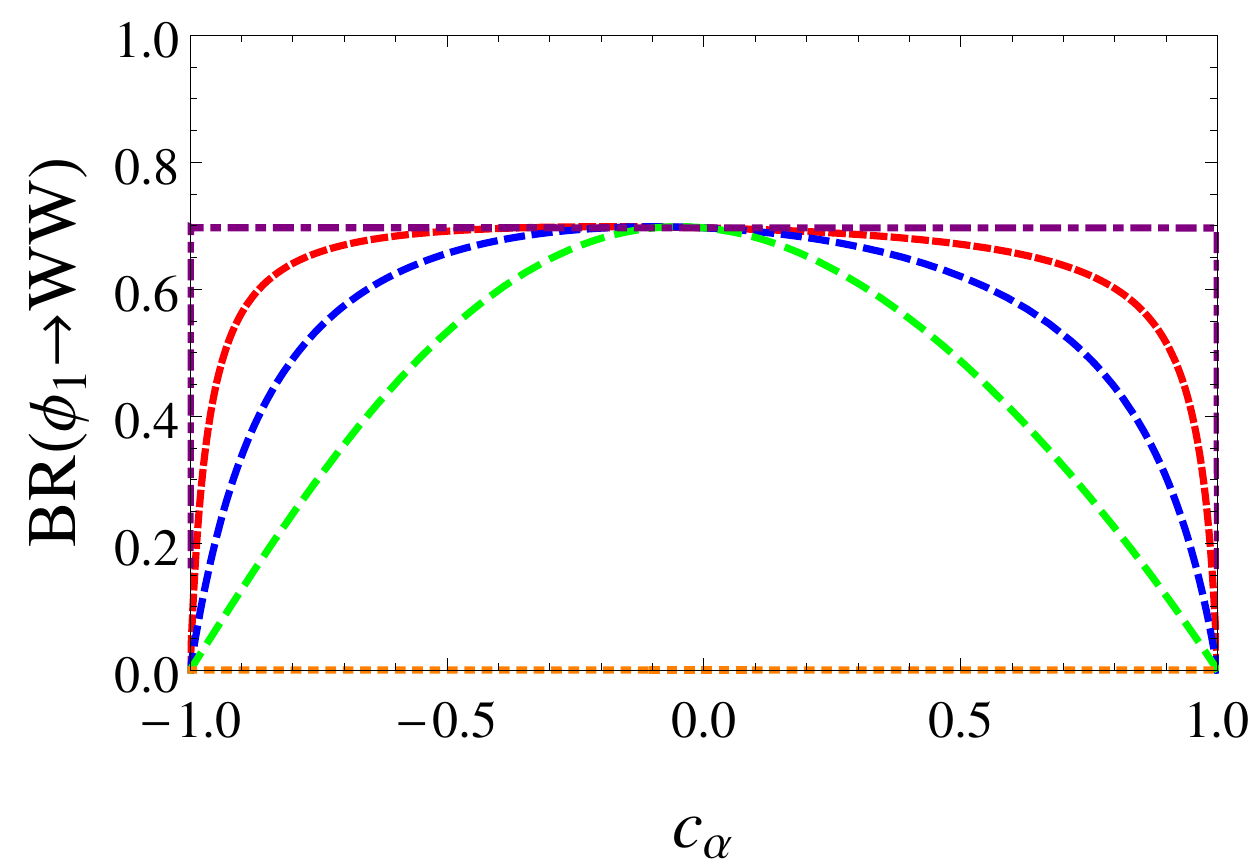}}\\
\subfloat[]{\includegraphics[height=3.5cm,width=3.8cm]{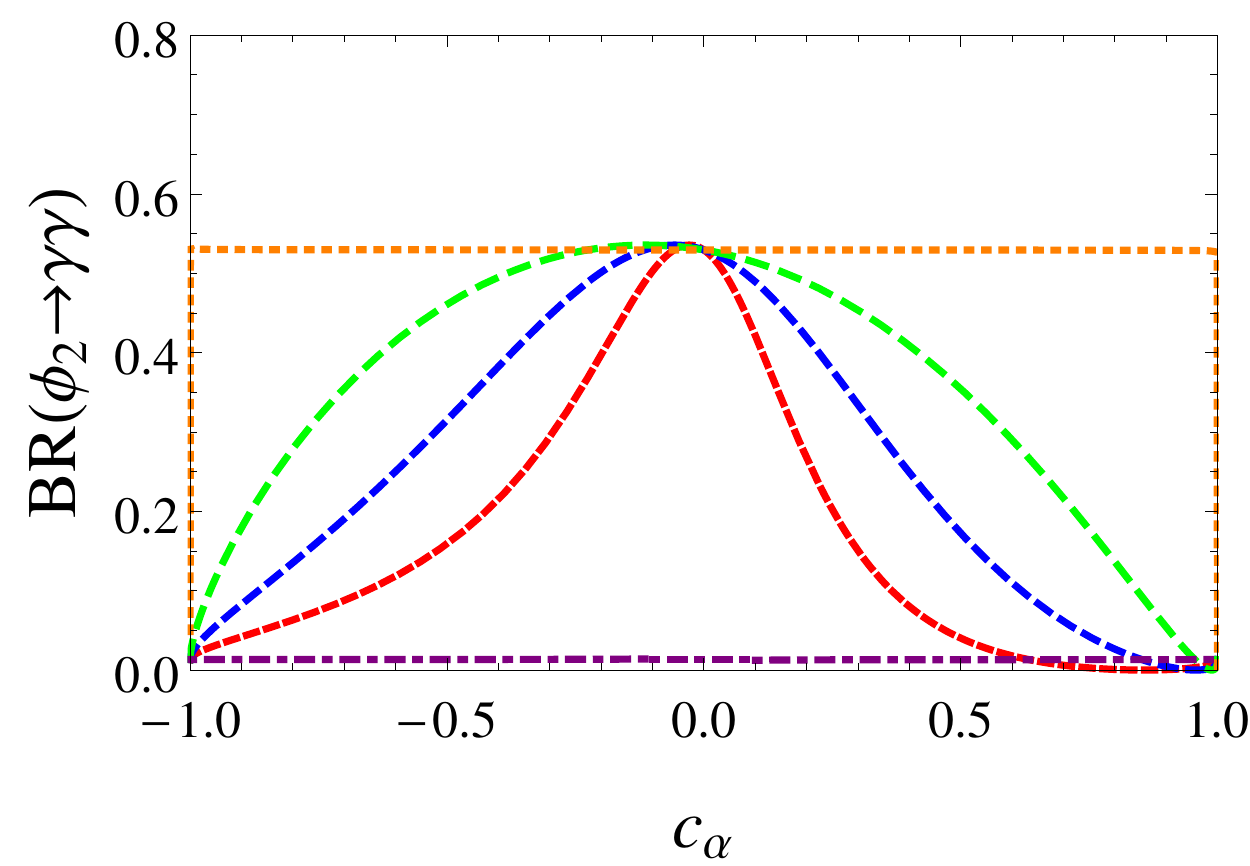}}\hspace{0.1cm}
\subfloat[]{\includegraphics[height=3.5cm,width=3.8cm]{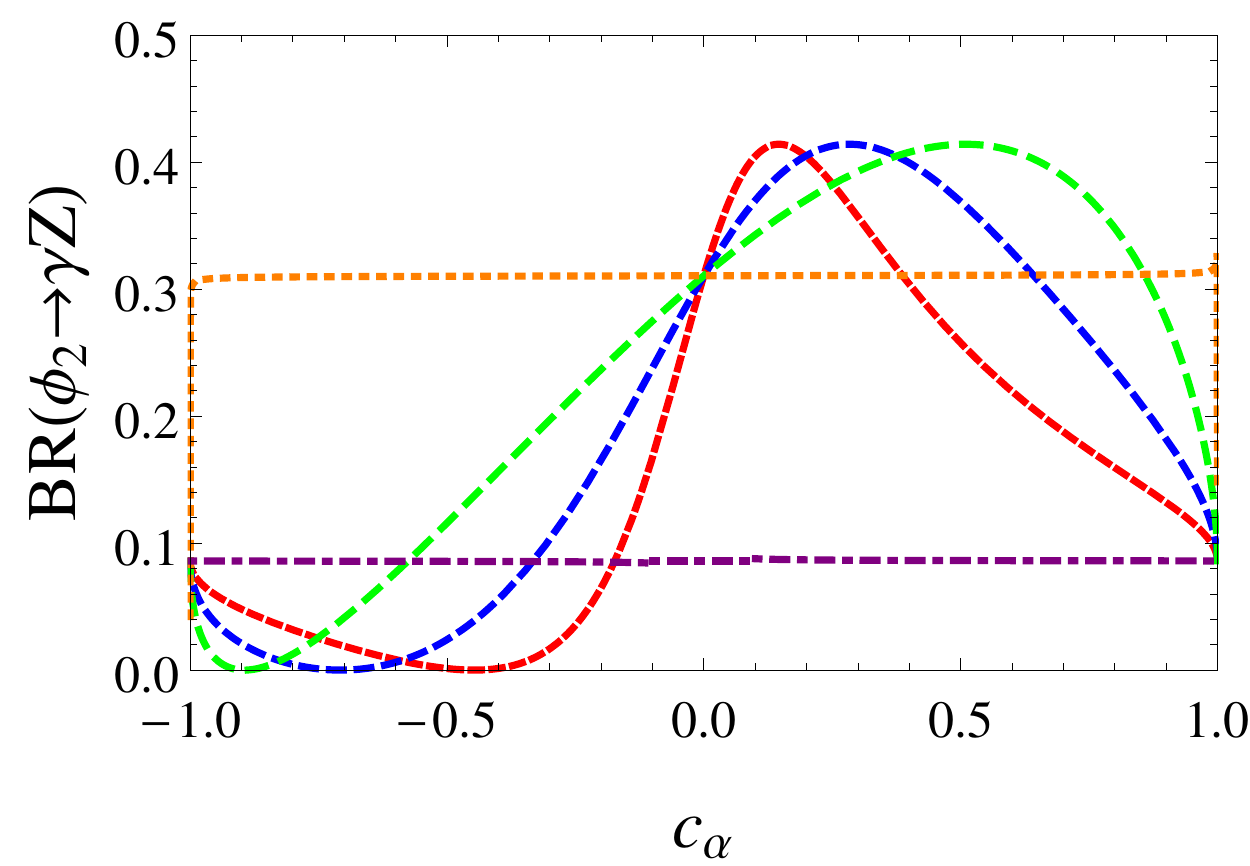}}\hspace{0.1cm}
\subfloat[]{\includegraphics[height=3.5cm,width=3.8cm]{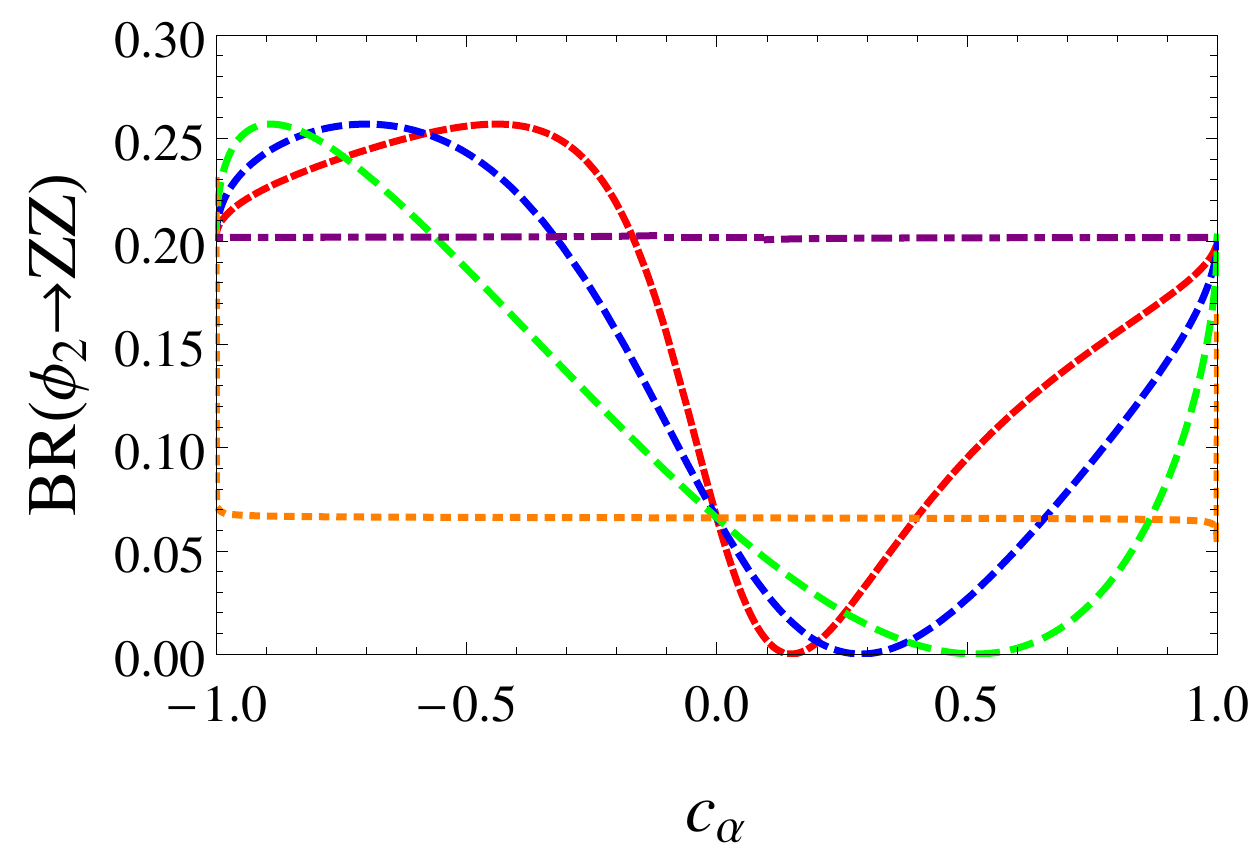}}\hspace{0.1cm}
\subfloat[]{\includegraphics[height=3.5cm,width=3.8cm]{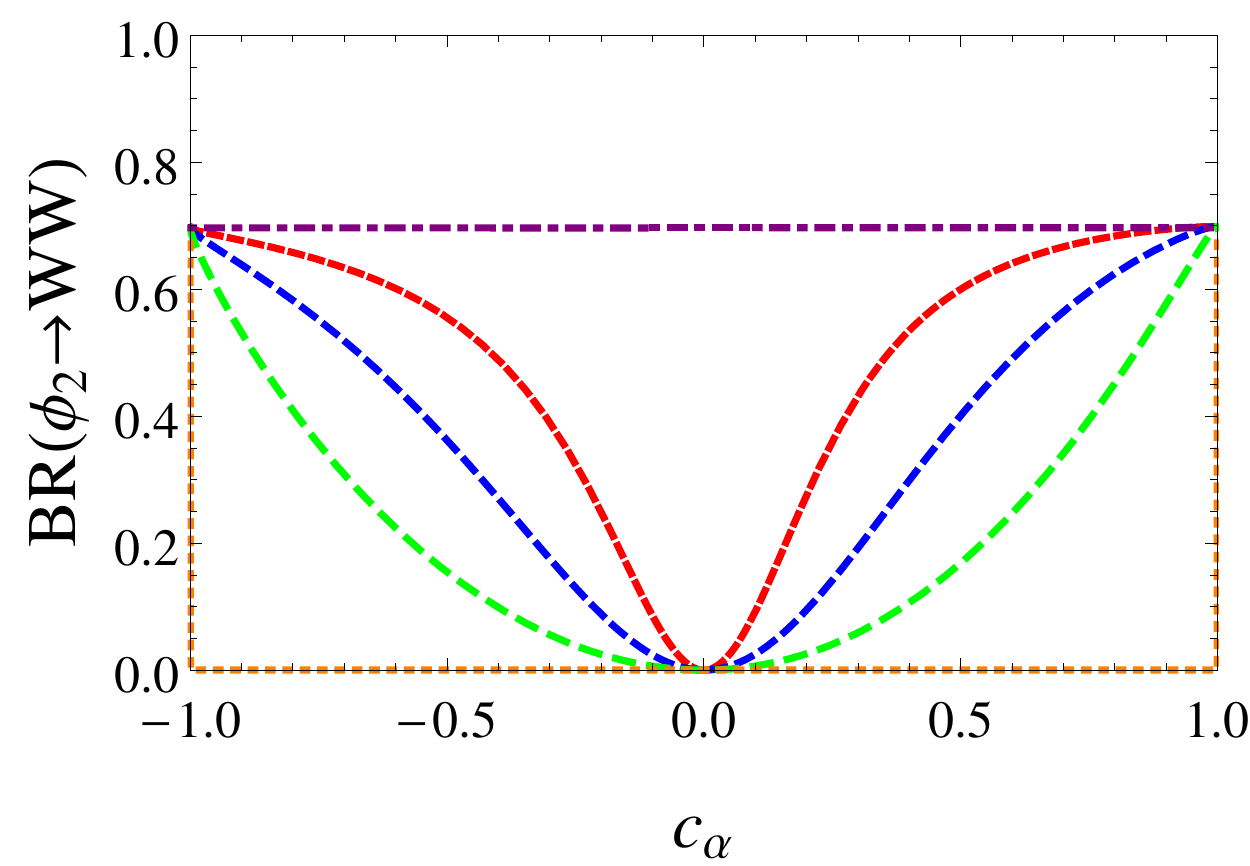}}\\
\includegraphics{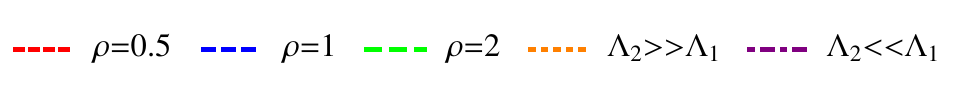}
\caption{Branching ratios of $\phi_1$ (a--d) and $\phi_2$ (e--h) into $\gm\gm,\gm Z,ZZ$ and $WW$, 
as functions of $\cos\alpha$ for $\rho=0.5,1,2$, $\Lm_2\gg\Lm_1$ (\emph{photophilic} limit) and $\Lm_2\ll\Lm_1$ (\emph{photophobic} limit). Here, $\phi_1$ and $\phi_2$ are the mass eigenstates in the two scalar theory, assuming 
$M_{\phi_1,\phi_2}=1$ TeV for these plots.}
\label{fig:BRphii}
\end{figure*}

We use the generic symbol $\mc{S}~=\lt\{S_0,\phi_1,\phi_2\rt\}$ for the new scalars (mass eigenstates) in the 1ST and 2ST models. These scalars dominantly decay to two SM gauge bosons, 
i.e., $\mc{S}\to\mc{V}_1\mc{V}_2$.
Interestingly, there is no $S_0\gm Z$ coupling in 1ST, and therefore $S_0\to \gm Z$
decay is not possible. In 2ST, however, all four two-body decay modes are possible. 
Below, we show the expressions for the partial widths of the scalars $\phi_i$ in the 2ST model, and similar
expressions can be readily obtained for $S_0$ in the 1ST by changing the corresponding decay couplings.
According to 
Eq.~\eqref{eq:effint}, the partial widths of the two-body decays of $\phi_i$ are given by
\begin{align}
\label{eq:pwformula}
\Gm_{\gm\gm}^i&=\frac{\lt(\kp_{\gm\gm}^i\rt)^2}{4\pi}M_{\phi_i}^3\nn\\
\Gm_{\gm Z}^i&=\frac{\lt(\kp_{\gm Z}^i\rt)^2}{8\pi}M_{\phi_i}^3\lt(1-\frac{M_Z^2}{M_{\phi_i}^2}\rt)^3 \nn\\
\Gm_{VV}^i&=\frac{3\lt(\kp_{VV}^i\rt)^2}{8\pi f_V}M_{\phi_i}^3\lt(1-\frac{4M_V^2}{M_{\phi_i}^2}\rt)^{\frac{5}{2}},
\end{align}
where $V=\lt\{Z,W\rt\}$ and $f_Z=1$ and $f_W=2$. There are also three-body decay modes 
possible and among them $\phi_i\to\gm ff$ (where $f$ is a charged fermion) mediated by an off-shell photon is
substantial. For instance, $\Gm_{\gm ff}^i/\Gm_{\gm\gm}^i\approx 0.18$ for $M_{\phi}=1$ TeV and
increases to 0.21 for $M_{\phi}=3$ TeV. We compute $\Gm_{\gm ff}^i$ analytically using the expressions
given in Appendix~\ref{app:3bd} and also numerically using \textsc{MadGraph}~\cite{Alwall:2014hca}.
In our analysis, the additional contributions of the three-body partial widths have been properly included in the total widths of the scalars.

The 1ST model is very economical and predictive as it has only the two free parameters 
$M_{S_0}$ and $\Lm_0$.
In Fig.~\ref{fig:BRphi}, we show the BRs of $S_0$ as functions of $M_{S_0}$. The BRs are independent of $\Lm_0$ and
almost independent of $M_{S_0}$: all partial widths are proportional to $M_{S_0}^3$ when $M_{S_0}\gg M_V$,
which makes the BRs (almost) independent of $M_{S_0}$. In the limit $M_{S_0}\gg M_V$, the two-body BRs are in the proportions
\begin{align}
\textrm{BR}_{\gm\gm}:\textrm{BR}_{ZZ}:\textrm{BR}_{WW}\approx 18\%:26\%:52\%.
\end{align}
Note that the $S_0\to\gm ff$ mode has a BR of about 4\%.

There are in total five free parameters in the 2ST, 
$M_{\phi_1}$, $M_{\phi_2}$, $\Lm_1$, $\Lm_2$, and $\al$. 
In general, the $\Lm_1$ and $\Lm_2$ scales can be different, and we
define a parameter $\rho$ as $\Lm_2=\rho\Lm_1$. In Fig.~\ref{fig:BRphii}, we show the BRs of $\phi_i$ as functions of 
$c_\al\equiv\cos\al$ for five cases, $\rho=0.5,1,2$, $\Lm_2\gg\Lm_1$, and $\Lm_2\ll\Lm_1$.
For both $\phi_1$ and $\phi_2$, the largest BR to $\gm\gm$ is about 55\% for all $c_\al$ in the limit 
$\Lm_2\gg\Lm_1$, which we call the photophilic limit. Here, the $WW$ mode disappears.
In the other extreme limit, $\Lm_2\ll\Lm_1$, the $\gm\gm$ mode disappears and the $WW$ BR becomes the largest,
about 70\% for all $c_\al$. We refer to this limit as the photophobic limit.
The other two modes $\gm Z$ and $ZZ$ are always present except for specific $c_\al$ values depending
on the $\rho$ values.  

\subsection{Production channels at the LHC}

\begin{figure}
\includegraphics[height=6.2cm,width=7.5cm]{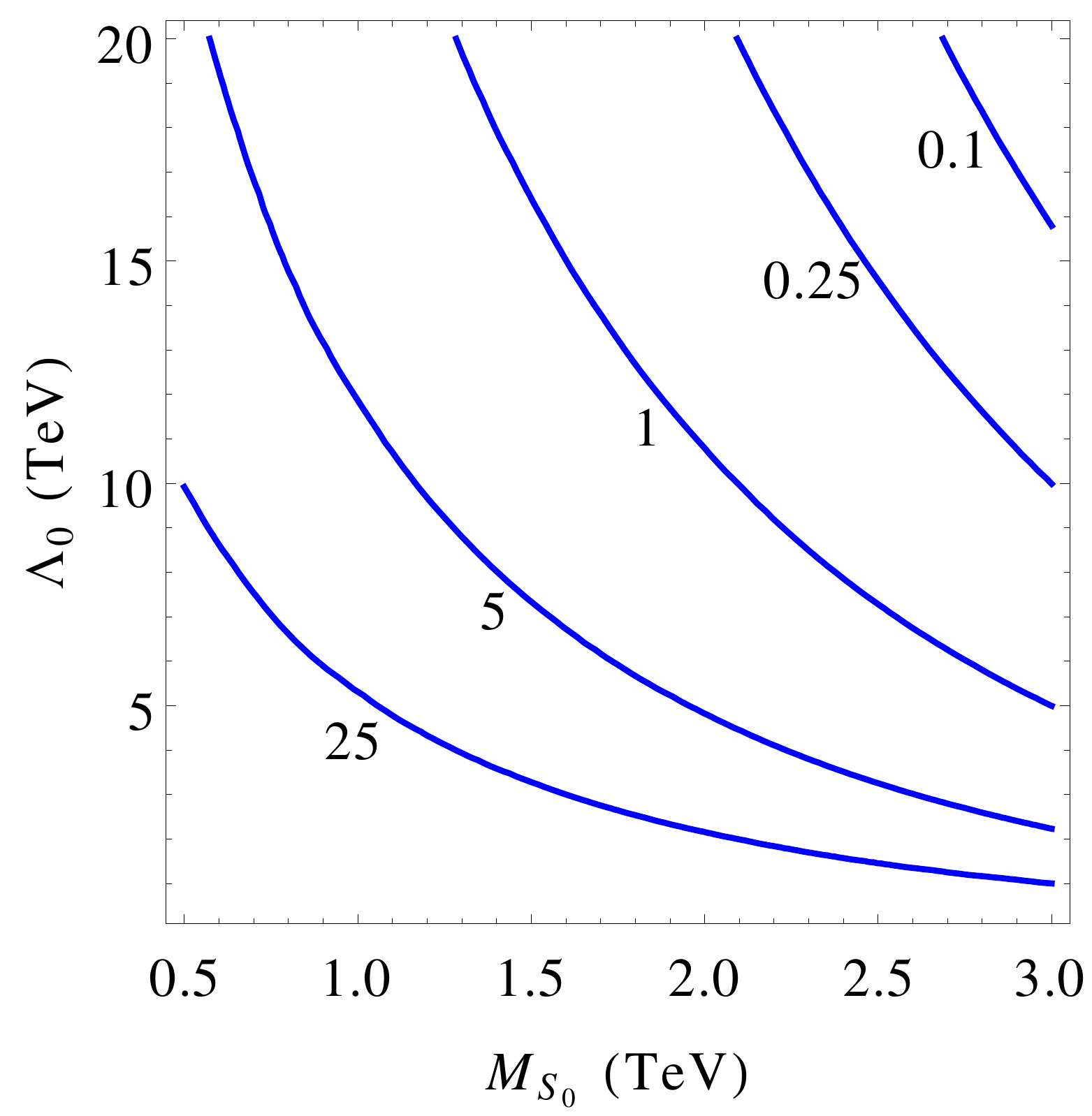}
\caption{Contours of production cross sections (in fb) of $S_0$ in the one scalar theory in the $M_{S_0}-\Lm_0$ plane for the 13 TeV LHC.}
\label{fig:CSphi}
\end{figure}

The scalars in the 1ST and 2ST models can be produced from $\mc{V}_1\mc{V}_2$ fusion processes, while the scalar $S_3$ associated with the strong coupling variation can be produced from $gg$ fusion.
In Fig.~\ref{fig:CSphi}, we show the total production cross section contours of $S_0$ in the 1ST model in the $M_{S_0}-\Lm_0$ plane 
at the 13 TeV LHC. The total cross section includes the contributions from the $\gm\gm,ZZ$ and $WW$ fusion 
production processes. We include photons as effective ``partons'' inside the proton. Therefore, the $\gm\gm$ initiated processes
can be directly generated in \textsc{MadGraph} (using the syntax \texttt{generate a a > S}). However, 
in case of $ZZ$ or $WW$ fusion, the initial $W$ and $Z$ come from quarks. 
Therefore, in this case, the scalars are produced in association with at least two additional non-QCD jets.
Similarly, in the 2ST model, the $\phi_i$ are produced in association with at least one non-QCD jet in case of $\gm Z$ initiated production.

\begin{figure*}
\subfloat[]{\includegraphics[height=6cm,width=7.5cm]{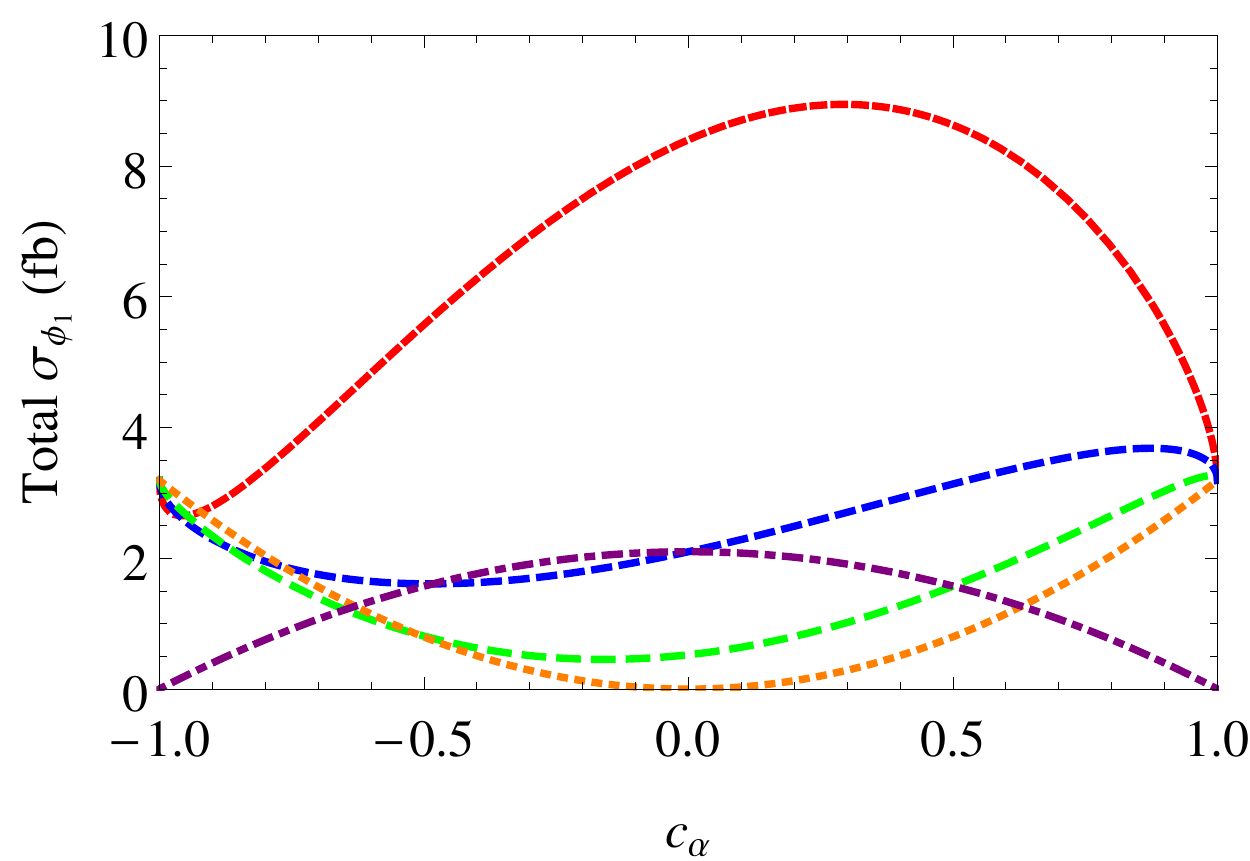}\label{fig:CSphi1}}
\subfloat[]{\includegraphics[height=6cm,width=7.5cm]{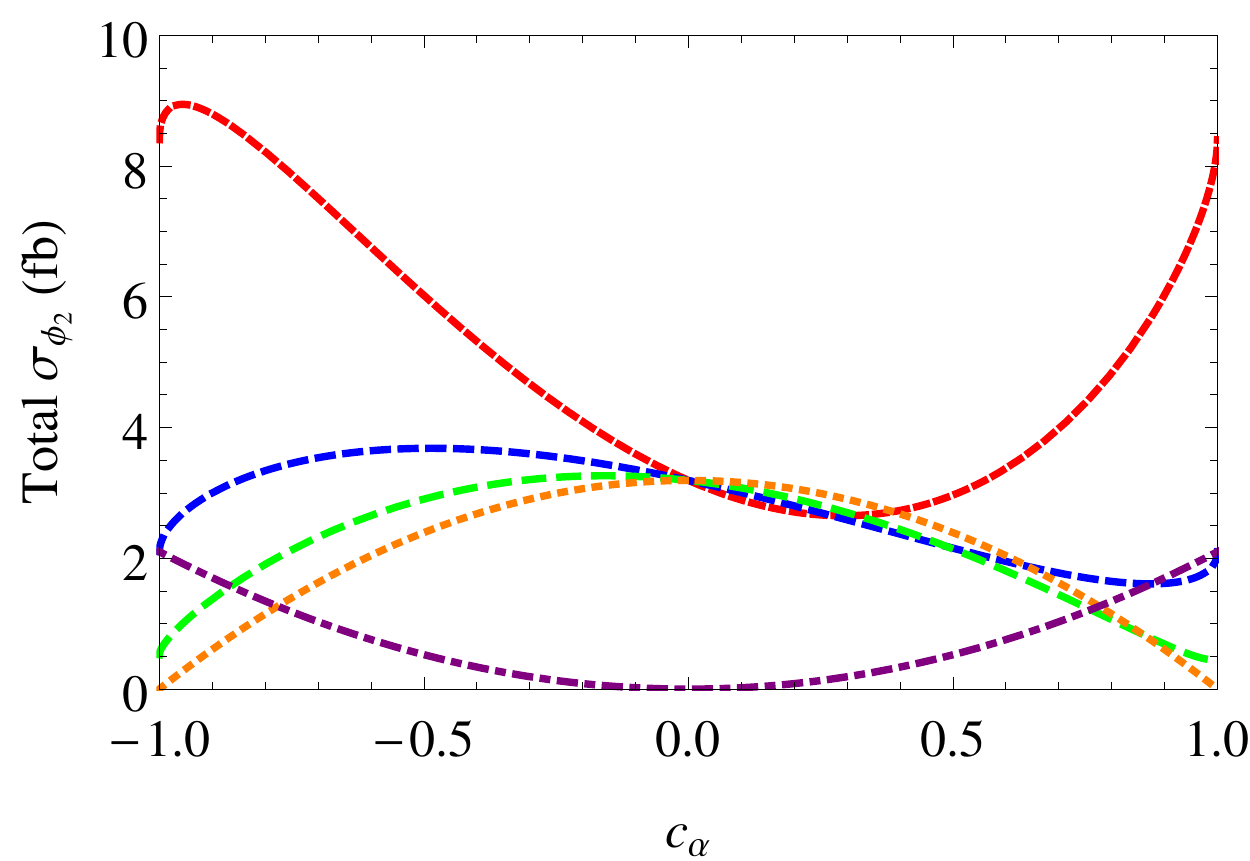}\label{fig:CSphi2}}\\
\includegraphics{figures/plotlegend}
\caption{Total production cross sections (in fb) of $\phi_i$ in the two scalar theory, as functions of $c_\al$ for
$M_{\phi_1}=1$ TeV with $\Lm_1=10$ TeV at the 13 TeV LHC. We present cross sections for $\rho=0.5,1$, and
2. In the photophilic limit, $\Lm_2\gg \Lm_1$ we set $\Lm_1=10$ TeV and $\Lm_2\to\infty$, whereas in the photophobic limit $\Lm_2\ll \Lm_1$, we set $\Lm_2=10$ TeV and $\Lm_1\to\infty$.}
\label{fig:CSphii}
\end{figure*}

\begin{figure}
\includegraphics[height=6.2cm,width=7.5cm]{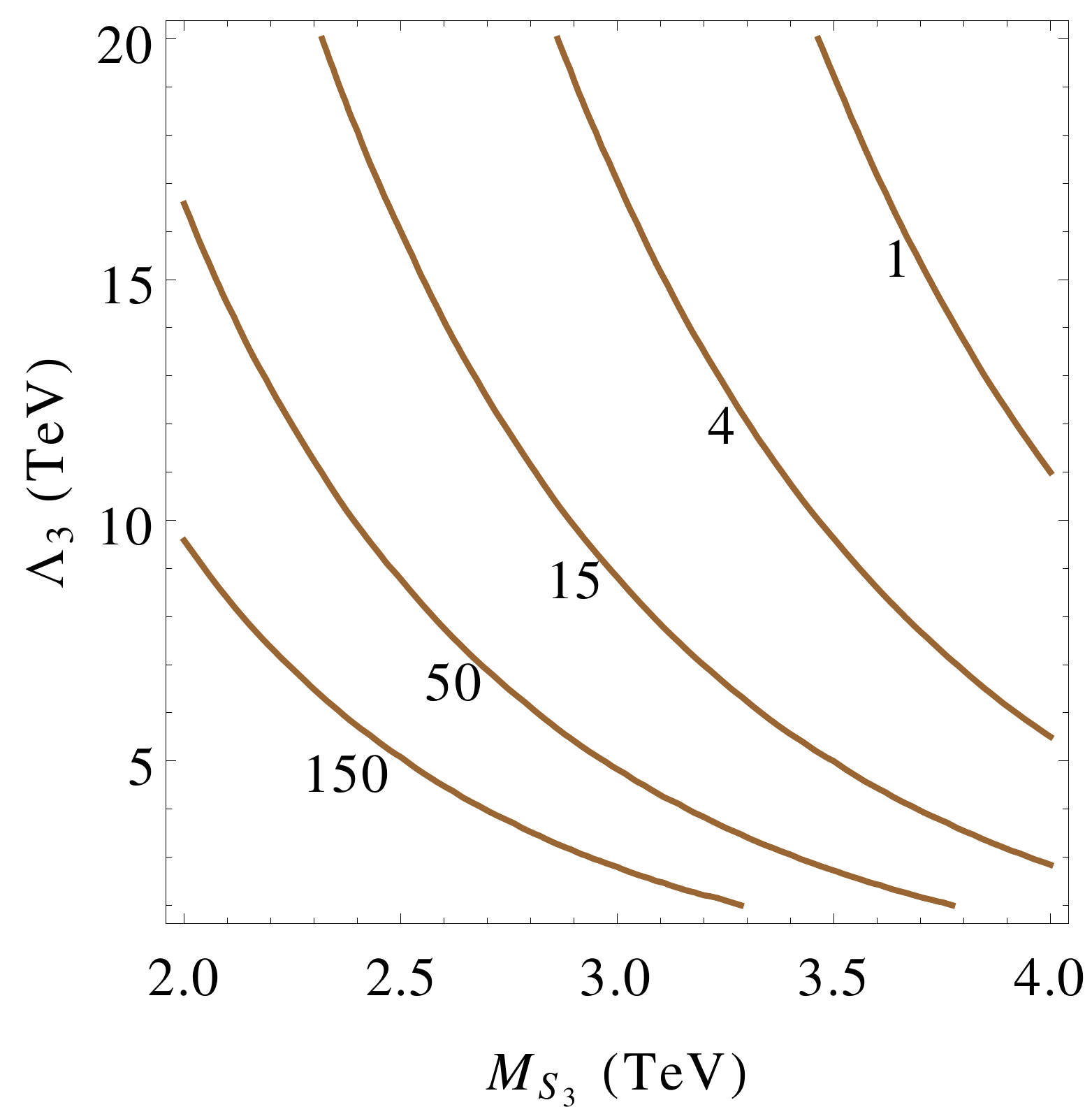}
\caption{Contours of production cross sections (in fb) of $S_3$ in the varying strong coupling theory in the $M_{S_3}-\Lm_3$ plane at the 13 TeV LHC.}
\label{fig:CSS3}
\end{figure}

The $\gm\gm,\gm Z$, and $ZZ$ production modes interfere among themselves. One has to take care to properly include the effects of these
interferences, which should be regarded as components of signal, in the total production cross sections of $\mc{S}$.
We calculate the total cross sections by combining the following processes, avoiding any double counting:
\begin{align}
\label{eq:process}
\textrm{Process 1:}~&\gm\gm\to \mc{S} \nn\\
\textrm{Process 2:}~&\gm\gm+\gm Z\to \mc{S}j \nn\\
\textrm{Process 3:}~&\gm\gm+\gm Z+ZZ+WW\to \mc{S}jj.
\end{align}
In process 1, both photons come from the proton, as explained above. In process 2, one of the photons or the $Z$ is radiated from a quark rather than coming from the proton, so there is one additional jet. In process 3, both incoming bosons are radiated from quarks, so this is a vector boson fusion process.

The $\gm\gm\to \mc{S}j$ channel in process 2 can also arise from the $\gm\gm\to \mc{S}$ with one initial or final state jet radiated,
but the interference of $\gm\gm$ and $\gm Z$ initiated processes can only be captured in process 2. 
Therefore, we take $\gm Z$ initiated and $(\gm\gm+\gm Z)$-interference contributions from process 2, while only the $\gm\gm$
initiated contribution is taken from process 1. Similarly,
$(\gm\gm+ZZ)$ and $(\gm Z+ZZ)$ interferences are captured in process 3 in addition to the pure $ZZ$ and $WW$ initiated contributions. Note that $WW$ initiated processes do not interfere with the others. See our previous paper \cite{Danielsson:2016nyy} for details of the calculation.

In Fig.~\ref{fig:CSphii}, we show the total production cross sections of $\phi_i$ in the 2ST model, computed using the 
method described above. In Fig.~\ref{fig:CSS3}, we show the total production cross section contours of $S_3$ in the VSC model in the $M_{S_3}-\Lm_3$ plane.
These cross sections include NLO QCD $K$-factors.
Not all $K$-factors for the processes in Eq.~\eqref{eq:process} are available in the literature. For the $\mc{V}_1\mc{V}_2$ fusion production modes, we assume a constant $K$-factor of 1.3~\cite{Arnold:2008rz}. 
The actual $K$-factors can differ slightly from this value and can also vary for different
masses of the scalars, but we have used a constant since it has very little effect on our results.
For the $gg\to S_3$ production, we compute the NNLO QCD $K$-factor using \textsc{Higlu} 
package~\cite{Spira:1995mt} as a function of $M_{S_3}$.

In the photophilic ($\Lm_2\gg\Lm_1$) limit, both $\phi_1$ and $\phi_2$ are produced mostly from
$\gm\gm$ fusion. Since in this limit all $\kp_{\mc{V}_1\mc{V}_2}^1 \propto c_\al$, the total cross section 
$\sg(pp\to\phi_1)\to 0$ for $c_\al\to 0$, whereas $\sg(pp\to\phi_2)$ is the largest for $c_\al=0$.

\section{Exclusion limits}
\label{sec:exclu}

Depending on the parameter values of the model, the scalars can be substantially produced in
the different $\mc{V}_1\mc{V}_2\to\mc{S}$ fusion channels, followed by decays $\mc{S}\to\mc{V}_1\mc{V}_2$.
Therefore, $\gm\gm$~\cite{Aaboud:2017yyg,Sirunyan:2018wnk}, $\gm Z$~\cite{Sirunyan:2017hsb,Aaboud:2018fgi}, and
$VV$~\cite{Aaboud:2018bun,CMS:2019sgt} ($ZZ$ and $WW$ with different decays) resonance searches at the LHC can be used to constrain the parameter space for the scalars associated with the EW sector, and 
dijet resonance search data~\cite{ATLAS:2019bov,CMS:2018wxx} can be used to constrain the
parameter space of the VSC theory. These searches set upper limits
on $\sg\times{\rm BR}$ for $s$-channel resonances of different spin hypotheses. 
By recasting these upper limits, we can derive exclusion limits on our model parameters. Various production modes of a heavy scalar and its decay to different final states have been discussed in
Refs.~\cite{vonBuddenbrock:2016rmr,Mandal:2016bhl}. 

\begin{figure}[b!]
\includegraphics[height=6.0cm,width=7.5cm]{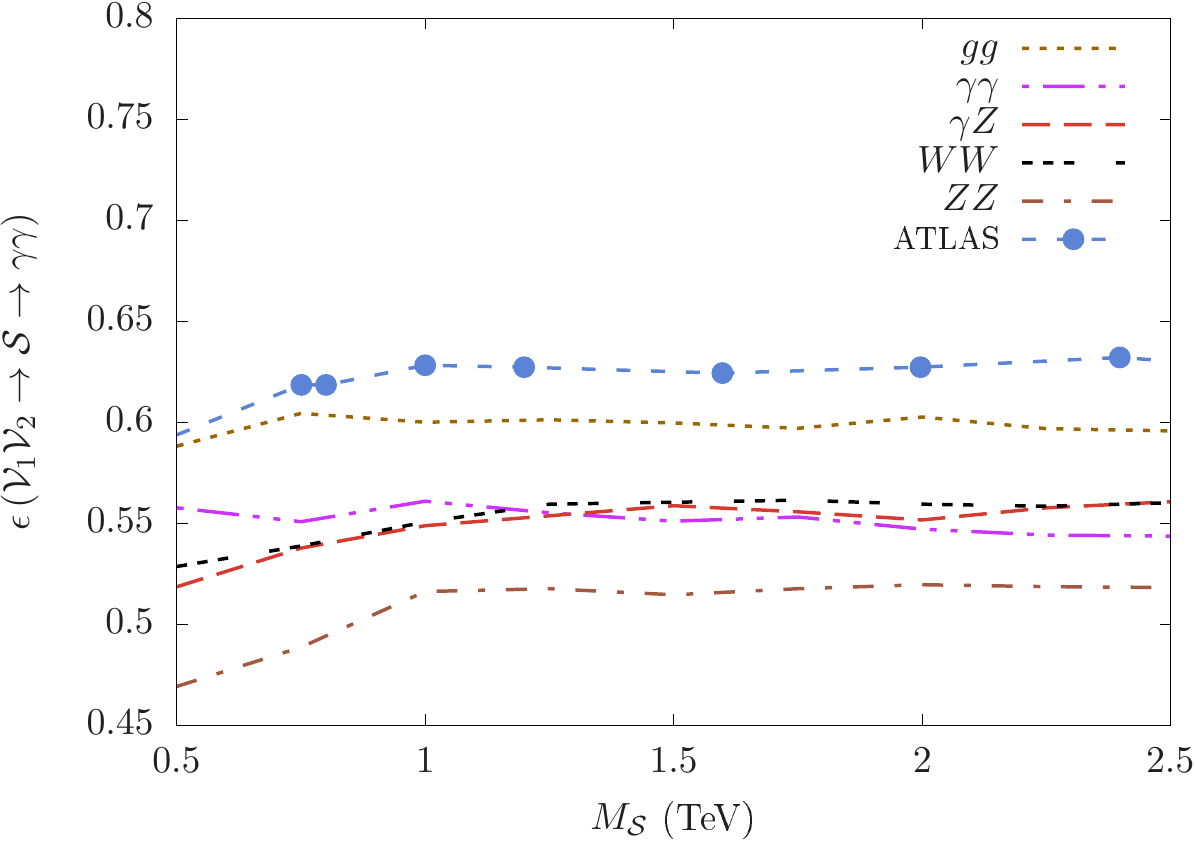}
\caption{Selection cut efficiencies for the spin-0 $\gm\gm$ resonance produced in
$\mc{V}_1\mc{V}_2$ fusion as functions of the resonance mass $M_{\mc{S}}$ for the 
fiducial selection cuts used in Ref.~\cite{Aaboud:2017yyg}. The cut efficiency obtained by the 
ATLAS Collaboration is shown by blue dots for a spin-0 resonance produced through the $gg$ fusion.}
\label{fig:CEATLASyy}
\end{figure}
\begin{figure*}
\subfloat[$\gm\gm$]{\includegraphics[height=0.33\textwidth]{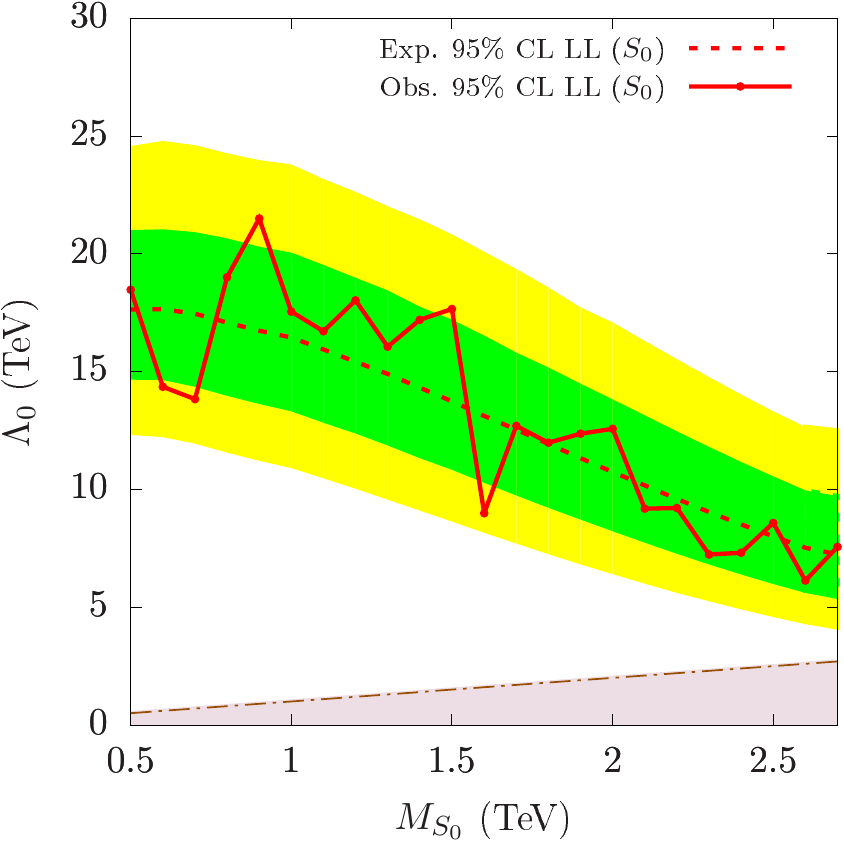}\label{fig:Mphi0LM0yy}}
\subfloat[$ZZ$]{\includegraphics[height=0.33\textwidth]{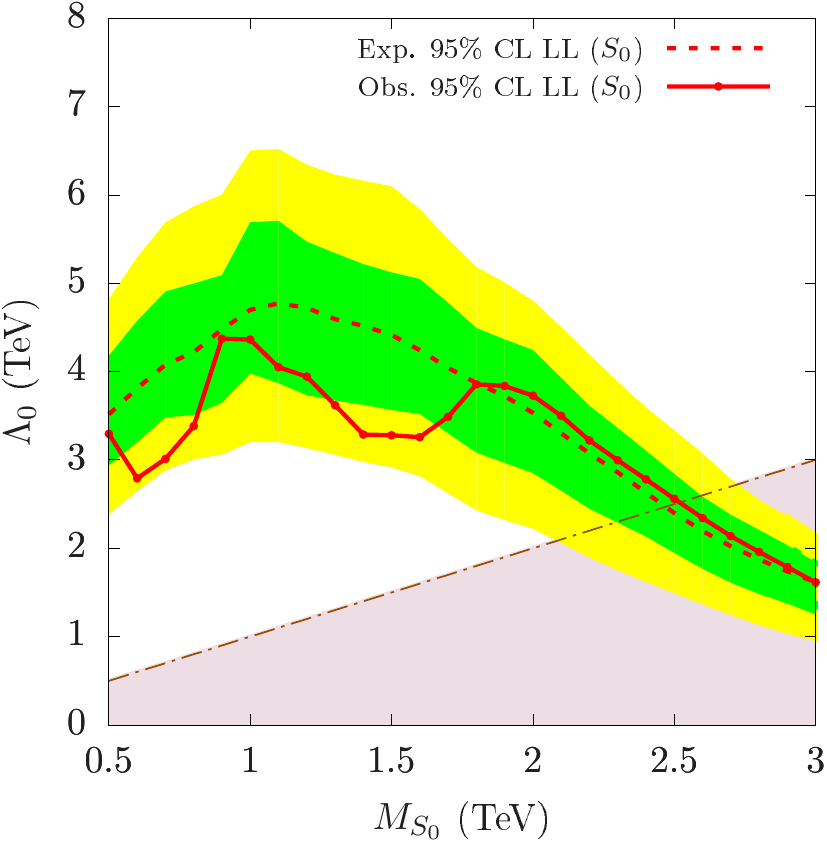}\label{fig:Mphi0LM0zz}}
\subfloat[$WW$]{\includegraphics[height=0.33\textwidth]{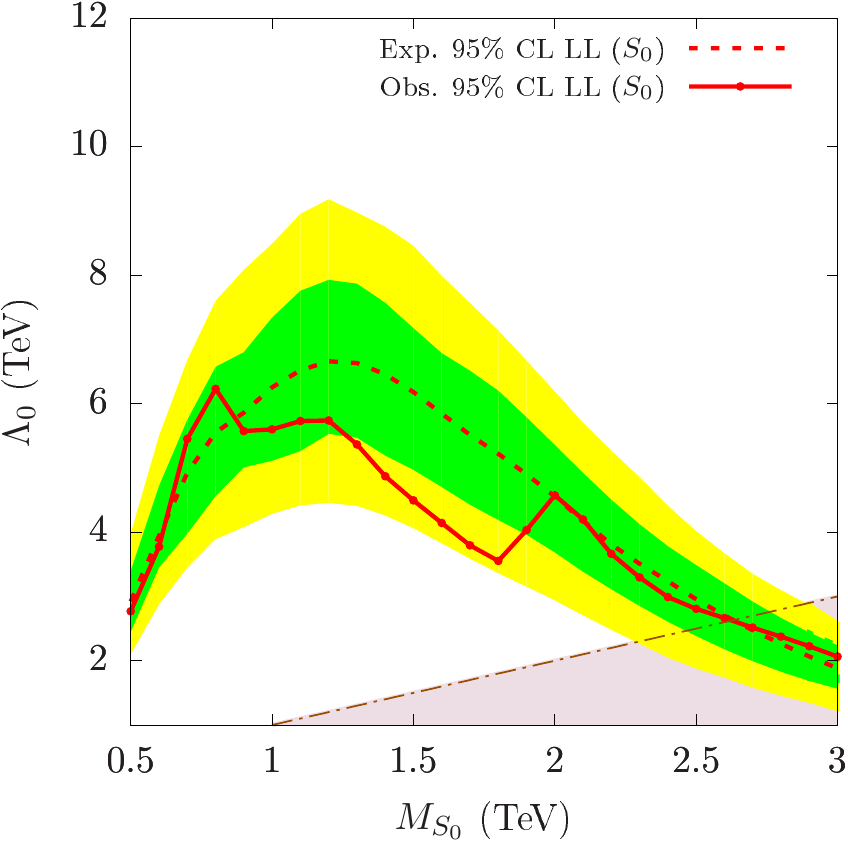}\label{fig:Mphi0LM0ww}}
\caption{The lower limits on $\Lm_0$ as functions of $M_{S_0}$ in the one-scalar theory, derived by recasting the upper limits on
$\sg\times{\rm BR}$ set by the ATLAS Collaboration in (a) $\gm\gm$~\cite{Aaboud:2017yyg}, (b) $ZZ$~\cite{Aaboud:2018bun}, and (c) $WW$~\cite{Aaboud:2018bun}. The solid and dashed curves are the observed and expected exclusion limits, respectively. 
The $1\sg$ and $2\sg$ uncertainties associated with the expected limits are represented by green and yellow bands. The regions
above the solid curves are allowed by the experiments. The light-colored regions in the background represent
$\Lm_0 < M_{S_0}$ where our effective field theory model is no longer valid. However, this validity condition is not strict and
is only shown for illustration.}
\label{fig:Mphi0LM0}
\end{figure*}
\begin{figure*}
\subfloat[$\gm\gm$]{\includegraphics[height=6.2cm,width=7.5cm]{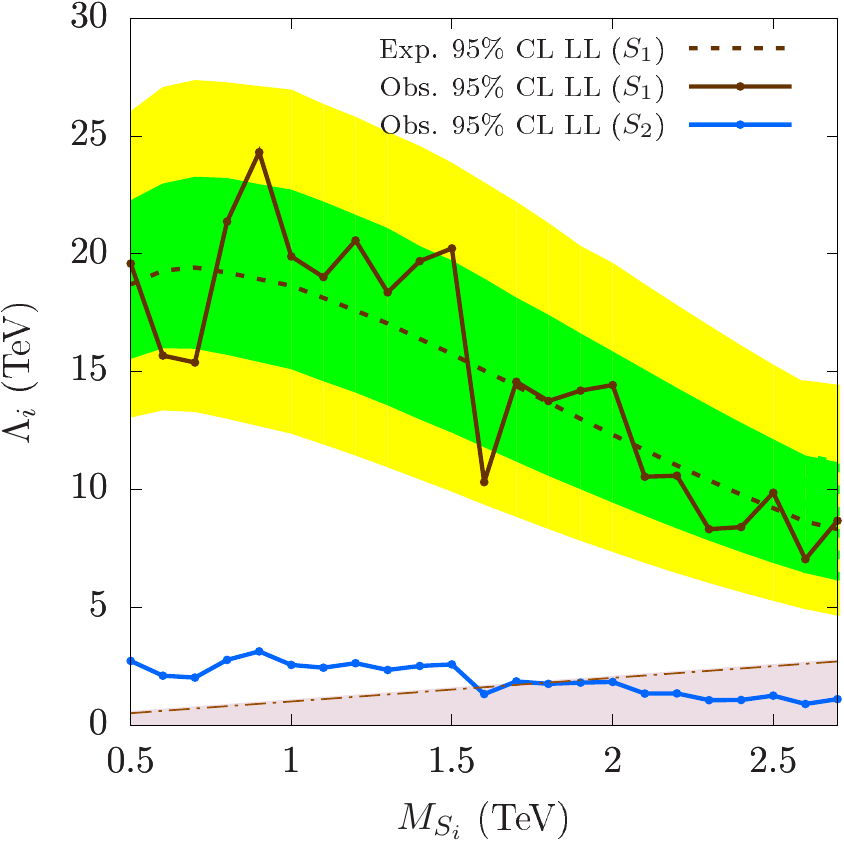}\label{fig:MphiiLMiyy}}\hspace{0.5cm}
\subfloat[$\gm Z$]{\includegraphics[height=6.2cm,width=7.5cm]{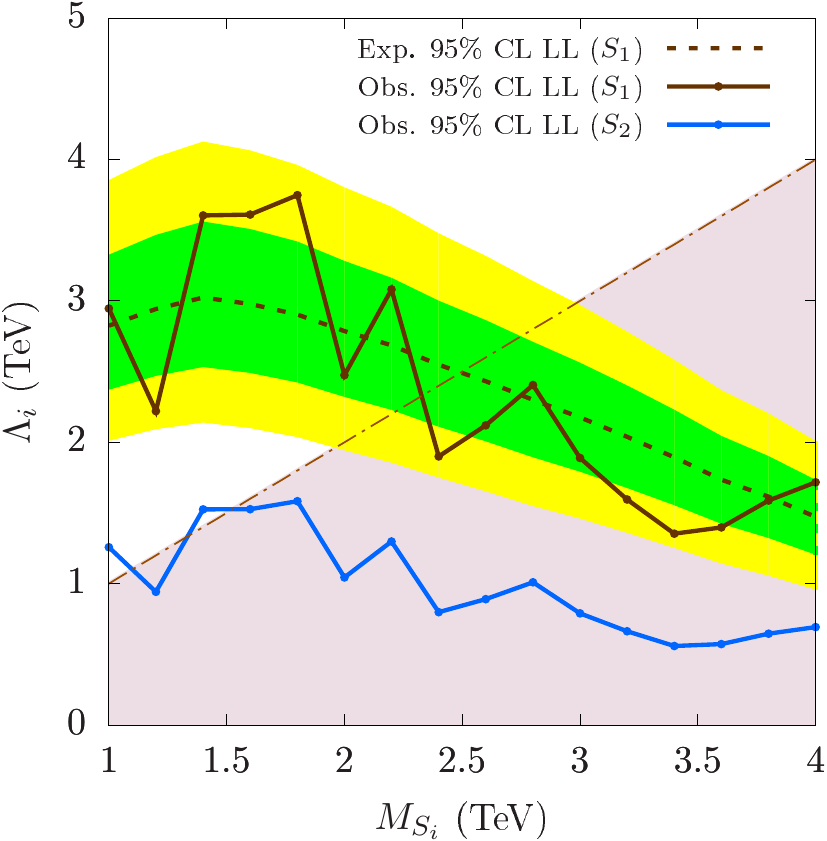}\label{fig:MphiiLMiyz}}\\
\subfloat[$ZZ$]{\includegraphics[height=6.2cm,width=7.5cm]{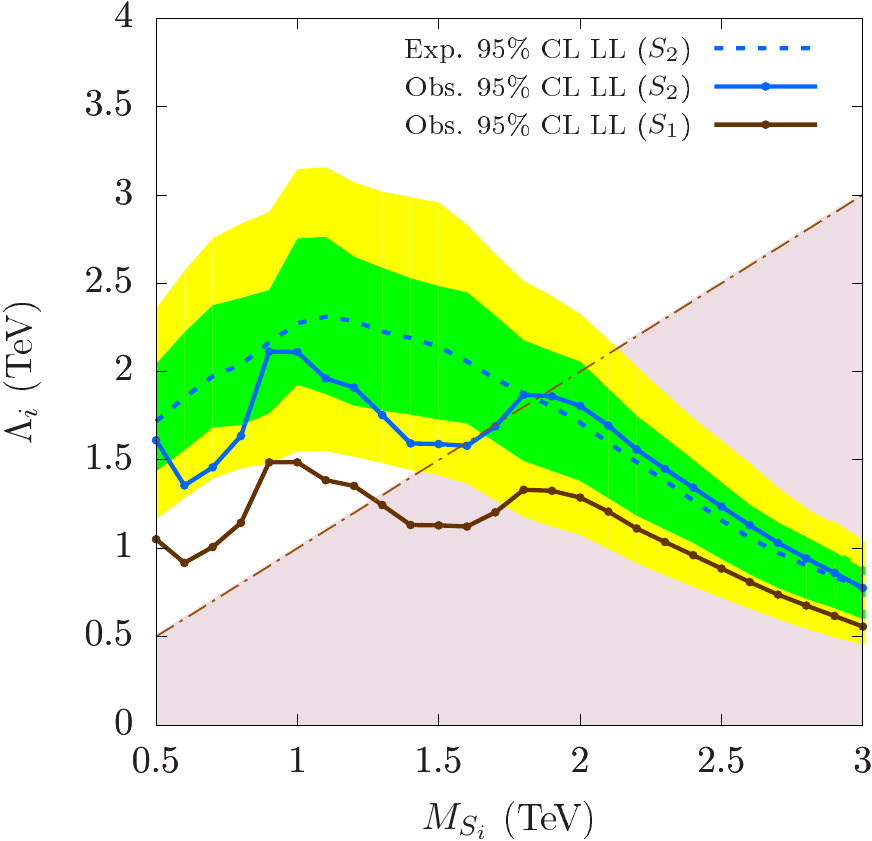}\label{fig:MphiiLMizz}}\hspace{0.5cm}
\subfloat[$WW$]{\includegraphics[height=6.2cm,width=7.5cm]{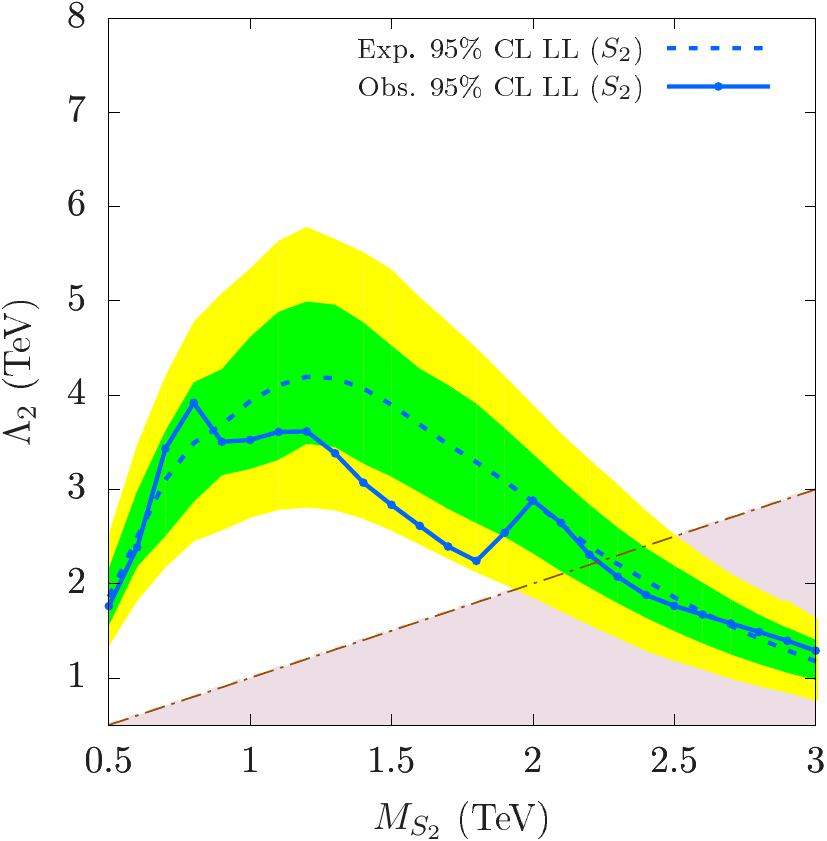}\label{fig:MphiiLMiww}}
\caption{Same as Fig.~\ref{fig:Mphi0LM0} for the two-scalar theory. Here, we present the limits in the no-mixing scenario ($c_\al=1$) 
for $\Lm_2\gg\Lm_1$ 
($S_1$ in (a)-(c)) and $\Lm_2\ll\Lm_1$ 
($S_2$ in (a)-(d)).}
\label{fig:MphiiLMi}
\end{figure*}

In general, these searches are
optimized for a resonance produced through $gg$ fusion.
For a specific final state, the selection cut efficiencies can be different for different production mechanisms.
Moreover, these efficiencies can vary for different spins of the resonances and also for different widths of the resonances.
Therefore, when deriving exclusion limits on the parameters by recasting upper limits on cross sections from 
various experiments, one must take care of the different cut efficiencies for different signal processes contributing
to the same final state. 
For this, we use the following relation from Refs.~\cite{Danielsson:2016nyy,Mandal:2016bhl}:
\be
\label{eq:Ns}
\mc{N}_s =(\sg\cdot {\rm BR})_s \times\ep_s\times L=\sum_i (\sg\cdot {\rm BR})_i\times\ep_i\times L\ ,
\ee
where the upper limit on the number of signal events is represented by $\mc{N}_s$. This can be expressed as the product
of the signal cross section (for a particular production mechanism), branching ratio (for a particular decay channel considered in the analysis), 
the corresponding selection cut efficiency $\ep_s$, and the integrated luminosity $L$.
In the case of different production mechanisms contributing to an experiment,
$\mc{N}_s=\sum_i (\sg\cdot {\rm BR})_i\times\ep_i\times L$ where $i$ runs over all the contributing
processes. 

We have checked
that the latest ATLAS and CMS results produce very similar exclusion limits and, in this paper, 
in the interest of simplicity, we use only the ATLAS results in our analysis.  
Below, we briefly discuss each of the searches used.

\medskip
\noindent
{\bf{ATLAS $\gm\gm$ resonance:~\cite{Aaboud:2017yyg}}} 
The ATLAS Collaboration
has performed searches for spin-0 and spin-2 $\gm\gm$ resonances produced in $gg$ fusion 
at the 13 TeV LHC with 37 fb$^{-1}$ integrated luminosity. We recast the fiducial $\sg\times{\rm BR}$ upper limit for a
spin-0 resonance with a 4 MeV width hypothesis (in the narrow width approximation). The definition of the fiducial
signal region can be found in~\cite{Aaboud:2017yyg}. We extract the observed and expected $\sg\times{\rm BR}$ upper limits (along with $1\sg$ and $2\sg$
uncertainty bands) from \textsc{HEPdata}~\cite{Maguire:2017ypu}.

In Fig.~\ref{fig:CEATLASyy}, we show how the selection cut efficiencies vary as functions of the $\gm\gm$ resonance mass for 
different production mechanisms. To validate our analysis code, we compare our fiducial selection cut efficiency for the $gg\to\mc{S}\to\gm\gm$ process with the corresponding ATLAS numbers and obtain close agreement within about $\sim 3\%$.
We find that the cut efficiencies for different production modes do not vary much ($\lesssim 15\%$).
To simplify our analysis, we neglect this small variation in cut efficiencies
while recasting this search. 

\medskip
\noindent
{\bf{ATLAS $\gm Z$ resonance:~\cite{Aaboud:2018fgi}}}
ATLAS has also performed searches for spin-0 and spin-2 $\gm Z$ resonances in $gg$ fusion with hadronic decays of the $Z$, with 36 fb$^{-1}$ integrated luminosity.
We recast the $\sg\times{\rm BR}$ upper limit derived for the spin-0 selection and, here too, we neglect the small variations in cut efficiencies.

\medskip
\noindent
{\bf{ATLAS $VV$ resonance:~\cite{Aaboud:2018bun}}}
The final searches relevant for the EW scalars are diboson (both $ZZ$ and $WW$) resonance searches from $gg$ and $VV$ fusion with $L=36$~fb$^{-1}$. As discussed above, the scalars $\mc{S}$ in our model can be produced in $VV$ fusion (or in the kinematically similar $\gm\gm$ or $\gm Z$ fusion processes)~\eqref{eq:process}. In~\cite{Aaboud:2018bun}, ATLAS presented combined upper limits on $\sg\times{\rm BR}$ for different bosonic and leptonic decay modes of $ZZ$ and $WW$, and we recast these combined limits for spin-0 resonance produced from $VV$ fusion.

\medskip
\noindent
{\bf{ATLAS $jj$ resonance:~\cite{ATLAS:2019bov}}}
This is the excited quark $q^{*}$ search that uses the $qg\to q^{*}\to qg$ process, with 139 fb$^{-1}$ 
integrated luminosity. In our model, we have a $gg$ resonance produced from the $gg\to S_3\to gg$ process. We have 
checked by employing the selection cuts used in~\cite{ATLAS:2019bov} that the cut efficiencies for these
two processes differ only by a few percent. This is expected as the spin information of the resonance is mostly lost in the
very boosted dijet system and they appear as two highly collimated jets in the detector. For simplicity, we neglect the
effect of cut efficiency in the derivation of exclusion limits.

\begin{figure}[tbh]
\includegraphics[height=6.2cm,width=7.5cm]{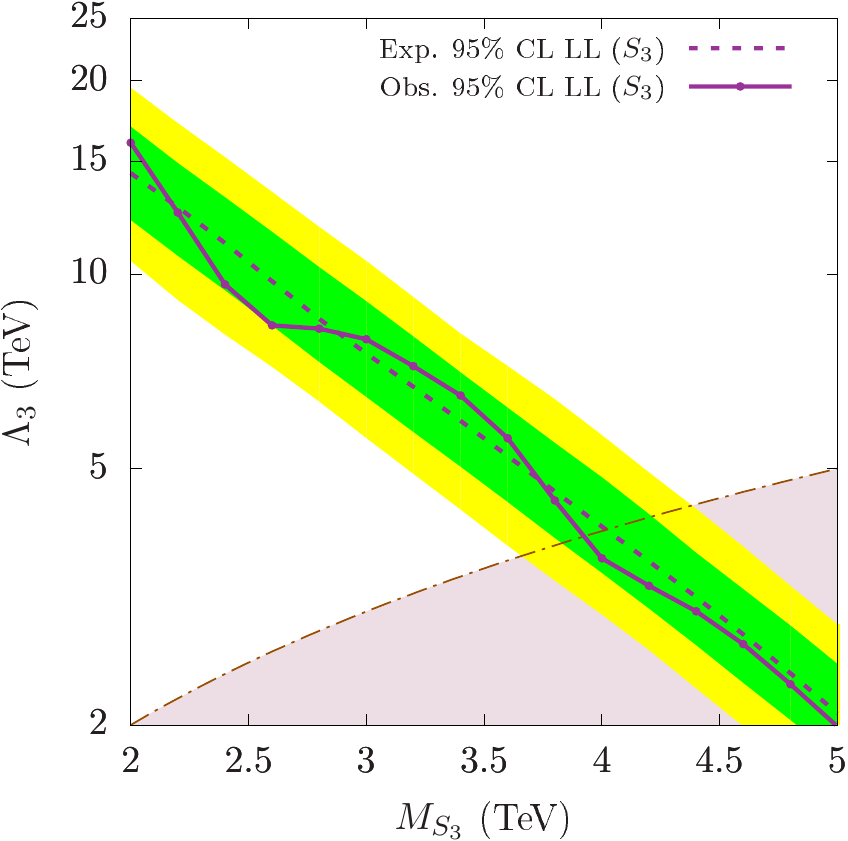}
\caption{The lower limit on $\Lm_3$ as a function of $M_{S_3}$ in the varying strong coupling theory, derived 
by recasting the upper limits on $\sg\cdot {\rm BR}$ as set by the ATLAS Collaboration in the $jj$~\cite{ATLAS:2019bov} resonance search. The meaning of the various colored regions is explained in the caption of Fig.~\ref{fig:Mphi0LM0}.}
\label{fig:Mphi3LM3}
\end{figure}

\begin{figure*}
\subfloat[$\gm\gm$]{\includegraphics[width=4.2cm]{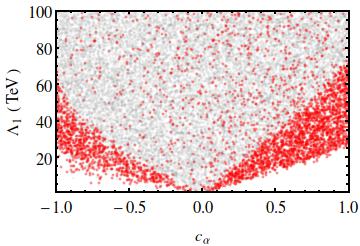}\label{fig:caLm1phi1yy}}
\subfloat[$\gm Z$]{\includegraphics[width=4.2cm]{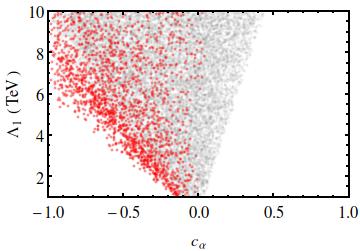}\label{fig:caLm1phi1yz}}
\subfloat[$ZZ$]{\includegraphics[width=4.2cm]{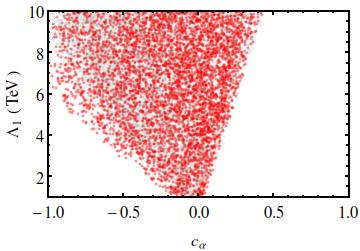}\label{fig:caLm1phi1zz}}
\subfloat[$WW$]{\includegraphics[width=4.2cm]{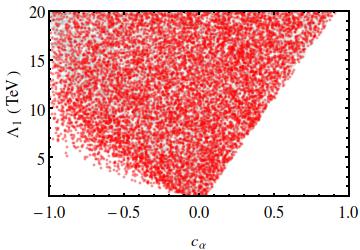}\label{fig:caLm1phi1ww}}\\
\subfloat[$\gm\gm$]{\includegraphics[width=4.2cm]{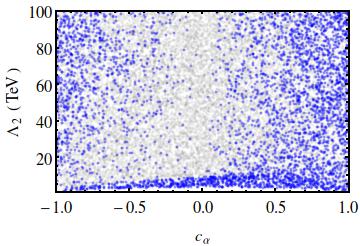}\label{fig:caLm2phi1yy}}
\subfloat[$\gm Z$]{\includegraphics[width=4.2cm]{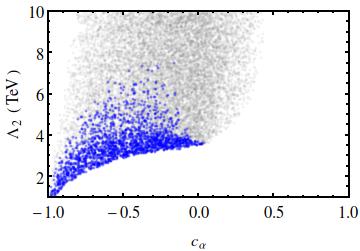}\label{fig:caLm2phi1yz}}
\subfloat[$ZZ$]{\includegraphics[width=4.2cm]{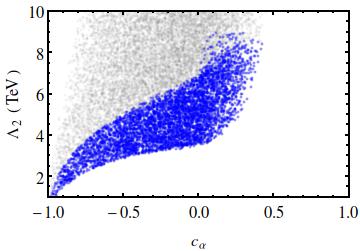}\label{fig:caLm2phi1zz}}
\subfloat[$WW$]{\includegraphics[width=4.2cm]{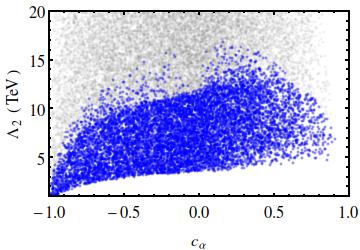}\label{fig:caLm2phi1ww}}\\
\subfloat[$\gm\gm$]{\includegraphics[width=4.2cm]{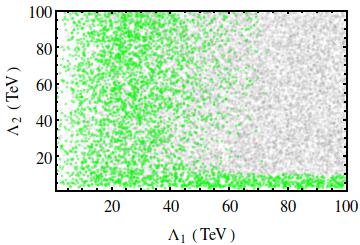}\label{fig:Lm1Lm2phi1yy}}
\subfloat[$\gm Z$]{\includegraphics[width=4.2cm]{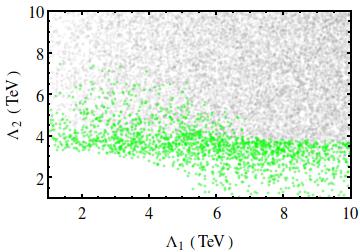}\label{fig:Lm1Lm2phi1yz}}
\subfloat[$ZZ$]{\includegraphics[width=4.2cm]{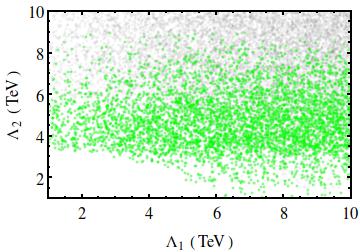}\label{fig:Lm1Lm2phi1zz}}
\subfloat[$WW$]{\includegraphics[width=4.2cm]{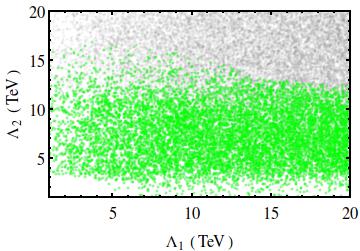}\label{fig:Lm1Lm2phi1ww}}
\caption{Scatterplots for $\phi_1$ in the $c_\al-\Lm_1$ (red dots), $c_\al-\Lm_2$ (blue dots) and 
$\Lm_1-\Lm_2$ (green dots) planes for the $\gm\gm$~((a),(e),(i)), $\gm Z$~((b),(f),(j)), $ZZ$~((c),(g),(k)), and $WW$~((d),(h),(l)) 
resonances. The colored regions can be reached with $>2\sg$ CL at the 13 TeV with $L=3000$ fb$^{-1}$. 
The white regions (no dots) in all figures represent the excluded regions, and the gray background dots represent the 
regions allowed by the current LHC data.}
\label{fig:highlumphi1}
\end{figure*}

\begin{figure*}
\subfloat[$\gm\gm$]{\includegraphics[width=4.2cm]{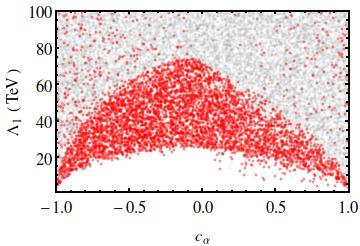}\label{fig:caLm1phi2yy}}
\subfloat[$\gm Z$]{\includegraphics[width=4.2cm]{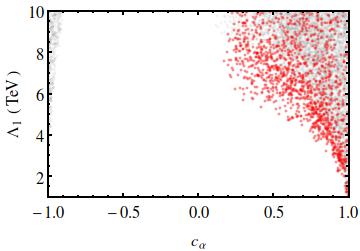}\label{fig:caLm1phi2yz}}
\subfloat[$ZZ$]{\includegraphics[width=4.2cm]{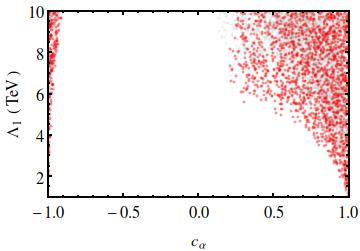}\label{fig:caLm1phi2zz}}
\subfloat[$WW$]{\includegraphics[width=4.2cm]{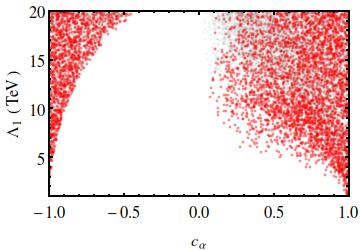}\label{fig:caLm1phi2ww}}\\
\subfloat[$\gm\gm$]{\includegraphics[width=4.2cm]{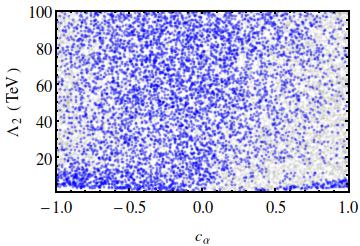}\label{fig:caLm2phi2yy}}
\subfloat[$\gm Z$]{\includegraphics[width=4.2cm]{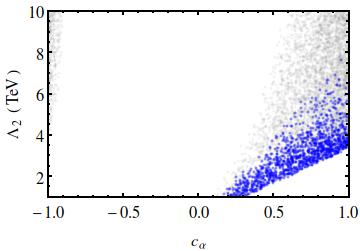}\label{fig:caLm2phi2yz}}
\subfloat[$ZZ$]{\includegraphics[width=4.2cm]{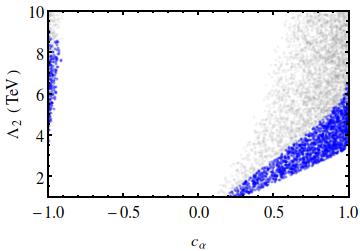}\label{fig:caLm2phi2zz}}
\subfloat[$WW$]{\includegraphics[width=4.2cm]{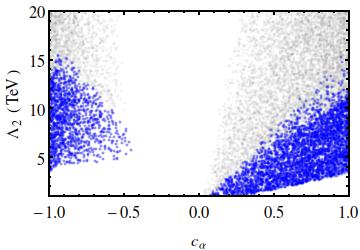}\label{fig:caLm2phi2ww}}\\
\subfloat[$\gm\gm$]{\includegraphics[width=4.2cm]{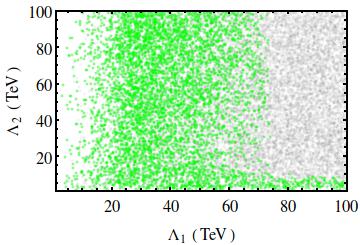}\label{fig:Lm1Lm2phi2yy}}
\subfloat[$\gm Z$]{\includegraphics[width=4.2cm]{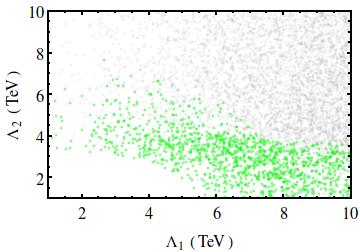}\label{fig:Lm1Lm2phi2yz}}
\subfloat[$ZZ$]{\includegraphics[width=4.2cm]{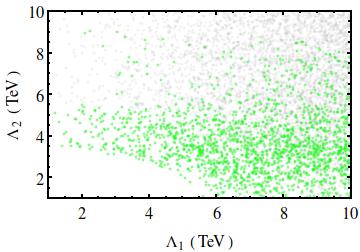}\label{fig:Lm1Lm2phi2zz}}
\subfloat[$WW$]{\includegraphics[width=4.2cm]{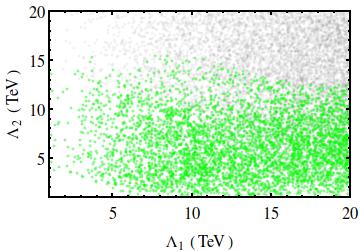}\label{fig:Lm1Lm2phi2ww}}
\caption{Same as Fig.~\ref{fig:highlumphi1} for $\phi_2$.}
\label{fig:highlumphi2}
\end{figure*}

\medskip
In Fig.~\ref{fig:Mphi0LM0}, we show the lower limits on $\Lm_0$ as functions of $M_{S_0}$ in the 1ST model 
from the $\gm\gm$, $ZZ$, and $WW$ resonance searches.
Although the $\gm\gm$ BR is about 18\% in 1ST, the strongest limit on $\Lm_0$ comes from the $\gm\gm$ channel.
For example, we obtain $\Lm_0\gtrsim 18$ TeV for $M_{S_0}\sim 1$ TeV from $\gm\gm$ data whereas $VV$ data can only
exclude $\Lm_0$ up to $\sim 5$ TeV for $M_{S_0}\sim 1$ TeV.

In Fig.~\ref{fig:MphiiLMi}, we show the lower limits on $\Lm_i$ in the 2ST model, as functions of $M_{S_i}$,  
from the $\mc{V}_1\mc{V}_2$ resonance searches.
In the photophilic limit ($\Lm_2\gg\Lm_1$), the $\gm\gm$ data can set strongest limits on $\Lm_1$, e.g., $\Lm_1\gtrsim 20$
TeV for $M_{S_1}\sim 1$ TeV. Here, we only show limits for the case when $S_1$ and $S_2$ are
mass eigenstates, i.e., with no mixing between them. Turning on $S_1\leftrightarrow S_2$ mixing can relax the
limits on $\Lm_1$ since the total production cross section $\sg_{\phi_1}$ is proportional to $c_\al^2$ in the
photophilic limit. In the photophobic limit ($\Lm_2\ll\Lm_1$) with no mixing, the $S_1$ decouples from the theory and 
the $\gm\gm$ data can only constrain $\Lm_2$ up to a few TeV. Overall, the limits on $\Lm_2$ are not very strong from the
latest LHC data, but $\Lm_1$ can be constrained strongly from the $\gm\gm$ data.

In Fig.~\ref{fig:Mphi3LM3}, we show the lower limit on $\Lm_3$ as a function of $M_{S_3}$ in the VSC. The 139 fb$^{-1}$ dijet resonance search data from ATLAS~\cite{ATLAS:2019bov} is
very powerful in constraining $\Lm_3$. For example, we estimate the lower limit on $\Lm_3$ to be about 16 TeV for $M_{S_3}\sim 2$ TeV.

\section{Reach at the HL-LHC}
\label{sec:reach}

In this section, we present the model parameter space that can be probed with more than $2\sg$ confidence level (CL) at the
HL-LHC, while at the same time satisfying the current data. We estimate the projected $2\sg$ CL upper limits of the
expected $\sg\times{\rm BR}$ by using a simple $\sqrt{L}$ scaling of the current expected values, i.e., 
\begin{align}
\label{eq:reachcond}
\sqrt{\frac{L_p}{L_f}}\lt(\sg\times {\rm BR}\rt)_{exp} < \lt(\sg\times {\rm BR}\rt)_{th} < \lt(\sg\times {\rm BR}\rt)_{obs},
\end{align}
where $\lt(\sg\times {\rm BR}\rt)_{obs}$ and $\lt(\sg\times {\rm BR}\rt)_{exp}$ are the observed and expected upper limits with 
present luminosity $L_p$, respectively, and the future luminosity is denoted by $L_f$. The theoretical $\lt(\sg\times {\rm BR}\rt)_{th}$ is a function of model parameters and can be calculated in a given theory. When the background is very small compared to the signal, the scaling goes as $L$. In reality, the actual scaling lies in between $\sqrt{L}$ and $L$, so 
using $\sqrt{L}$ scaling is more conservative and does not exaggerate the prospects.
At $L=3000$ fb$^{-1}$, the expected $\lt(\sg\times {\rm BR}\rt)_{exp}$ goes down by roughly 1 order of magnitude ($\sqrt{36/3000}\approx 0.11$). When translated to the projected lower limit on the new physics
scale $\Lm$, one finds that it goes
up by roughly a factor of 3. In case of the 1ST model, the obtained limits on $\Lm_0$ as shown in Fig.~\ref{fig:Mphi0LM0} go up by a factor
of 3 for $L=3000$ fb$^{-1}$. Similarly, for varying strong coupling theory, the limit on $\Lm_3$ in Fig.~\ref{fig:Mphi3LM3} 
goes up by a factor 2 for that luminosity. 

The situation gets more complicated in the case of the 2ST model, where we have five free parameters. To derive the simplified projected limits, we
perform a random scan over the three parameters $\Lm_1$, $\Lm_2$, and $c_\al$ in the regions
\begin{align}
\Lm_1,\Lm_2=\lt[1~\textrm{TeV},100~\textrm{TeV}\rt],~~c_\al=\lt[-1,1\rt],
\end{align}
 and fix the masses $M_{\phi_1},M_{\phi_2}=1$ TeV.
In Figs.~\ref{fig:highlumphi1} and \ref{fig:highlumphi2}, we show the outcomes of the random scan for $\phi_1$
and $\phi_2$, respectively. We present scatterplots in the planes of $c_\al-\Lm_1$ (red dots), $c_\al-\Lm_2$ (blue dots), and 
$\Lm_1-\Lm_2$ (green dots) 
for the four types of $\mc{V}_1\mc{V}_2$ resonances. These colored regions of the parameter space can be probed with at least
$2\sg$ CL at the HL-LHC with 3000 fb$^{-1}$ integrated luminosity. They are obtained by using the condition in
Eq.~\eqref{eq:reachcond} for a specific $\mc{V}_1\mc{V}_2$ resonance while satisfying others by the latest data, 
i.e., $\lt(\sg\times {\rm BR}\rt)_{th} < \lt(\sg\times {\rm BR}\rt)_{obs}$ only.
The white regions in all figures represent the presently excluded regions of the 
parameter space, and the gray background dots represent the regions which are allowed
by the current LHC data. 

For $\phi_1$, the $\gm\gm$ resonance can very effectively probe the $\Lm_1$ direction (for $c_\al$ away from zero)
but the $\Lm_2$ direction is mostly insensitive. On the other hand, the $WW$ resonance can be used to probe the $\Lm_2$
direction but not the $\Lm_1$ direction. For $\phi_2$, the $\gm\gm$ resonance can probe large $\Lm_1$ for $c_\al\sim 0$.
Since the $\gm\gm$ resonance is a very powerful channel to probe most of the parameter space of our models, it is
expected to observe these scalars in $\gm\gm$ resonance first before observing them in other channels. 

\section{Summary and conclusions}
\label{sec:conclu}

In this paper, we have investigated the phenomenology of heavy scalars associated with the variation of the gauge couplings 
of the SM. In our previous work~\cite{Danielsson:2016nyy}, we studied for the first time the collider phenomenology of a 
theory with a varying fine-structure constant (or in other words, a varying EM coupling $\widetilde{e}$) originally
proposed by Bekenstein~\cite{Bekenstein:1982eu}. We introduced a TeV-scale mass of the scalar associated
with the variation of $\widetilde{e}$, which therefore can be searched for at the LHC. In this work, we extend this
idea and explore for the first time the phenomenology of the three heavy scalars associated with the variation of the three gauge couplings
of the SM. Since these scalars are heavy, one has to go to high enough energy to excite them.
In the electroweak sector of the SM, the EM coupling $e$ appears only after EWSB. Therefore, it is more natural
to consider the simultaneous variations of the two gauge couplings $g_1$ and $g_2$ of the EW sector than to consider
the variation of $e$ only.
 
Above, we first discussed briefly the construction of the simplest and most economical model of a varying $\widetilde{e}$, with only two free parameters $M_{\phi}$ and $\Lm$. Following the same method, we then presented the construction of the model with varying
$g_1$ and $g_2$. We discussed two different models in this context. The one scalar theory has a single scalar $S_0$, which is 
responsible for the variation of both $g_1$ and $g_2$. This model is very economical with just two free parameters,
$M_{S_0}$ and $\Lm_0$, and hence is very predictive in terms of collider signatures. The two scalar theory, on the
other hand, has two different scalar fields, $S_1$ and $S_2$, which are responsible for the variations of $g_1$ and $g_2$, respectively. 
To make our analysis general, we consider the mixing of these two scalars among themselves and, therefore, introduce
a mixing angle $\alpha$ as one of the free parameters of the theory. This model is less predictive than the one scalar theory as it contains in
total five free parameters. However, this general setup is theoretically more pleasing and phenomenologically very rich, and one would expect many
interesting collider signatures, which we have discussed in detail. In addition, we also consider the model with 
varying strong coupling of the SM and the phenomenology of the associated scalar $S_3$. In order not to make our analysis more complex, we do not consider any mixing of $S_3$ with the other scalars.  

The scalars $S_1$ and $S_2$ can be produced in $\gm\gm,\gm Z,ZZ$, or $WW$ fusion at the LHC and can decay to
$\gm\gm,\gm Z,ZZ$, or $WW$. This can lead to various interesting signatures that we systematically discussed 
above. Depending on the free parameters of the theory, the branching ratios and production cross sections of these scalars 
show interesting patterns. We discussed these aspects by considering five benchmark scenarios, including two extreme limits
which we called the photophilic and photophobic limits. In the photophilic limit, the scalars have a strong affinity to couple to photons. In the photophobic limit, on the other hand, the scalars
mostly couple to $W$ bosons. The strengths of their couplings also depend on the mixing angle $\alpha$. We present
the branching ratios and production cross sections of these scalars as functions of $\cos\alpha$. 

We have derived exclusion limits of the model parameters by making use of the latest LHC $\gm\gm,\gm Z,ZZ,WW$, and $jj$
resonance search results. We found that the $\gm\gm$ search in particular is very powerful in constraining most of 
parameter region of the varying electroweak theory. For example, the $\gm\gm$ search sets lower limit on the
new physics scale roughly around 20 TeV for a scalar mass around 1 TeV. Using dijet resonance search data,
the scale $\Lm_3$ can be excluded up to 15 TeV for the $M_{S_3}\sim 2$ TeV. We also presented a simplified estimation of the sensitivity to our model parameter space at high luminosity LHC.  

In this paper, we limit ourselves to the varying gauge couplings of the SM. As discussed in the Introduction, this paper is related to several theoretical developments that may now be investigated with respect to observable signatures. In the same spirit, one can extend our work by promoting other dimensionless constants of the SM to scalar fields. For example, one may consider the variation of Yukawa couplings and the Higgs quartic coupling. The multiple scalars present in our framework might provide a strong first order electroweak phase transition, as required for baryogenesis. A quintessence potential, as introduced in \cite{Han:2018yrk} to address the vacuum stability of the Higgs potential, is of a similar structure to the potential and interactions that arise in our framework if we let the Higgs quartic coupling vary and promote the variation to a scalar field. These examples, and the first study presented in this paper, illustrate the future observational potential of the fundamental theoretical issue of whether what we now consider to be constants of nature have a deeper physical origin, as suggested in the considered extensions of the Standard Model.

\section*{Acknowledgments}
This work is supported by the Swedish Research Council under Grant No. 2015-04814. 
T.M. is grateful for support from The Royal Society of Arts and Sciences of Uppsala.

\section*{Appendices}
\appendix

\section{Derivation of the electromagnetic Lagrangian}
\label{app:derivation}
In this appendix, we will for completeness derive the Lagrangian for a varying EM coupling shown in Sec.~\ref{sec:model}, including the Bekenstein scalar and the constant coupling $e$. This is generalized to the other gauge groups in 
Sec.~\ref{sec:EWmodel}. We define the regular QED field strength tensor without the scalar field as $\fmn=\dmu A_\nu - \dnu A_\mu$, while the varying field strength tensor is given by $\hmn = {\ve^{-1}} \left[ \dmu (\ve A_\nu) - \dnu (\ve A_\mu) \right]$.
We will be assuming small fields, and write  $\ve \simeq 1+\varphi$, where $\varphi$ is defined in Sec.~\ref{sec:model}. In the following, we keep only terms linear in $\varphi$ or its derivatives.

We begin with the coupling of $\varphi$ to photons. Consider the kinetic term for the EM field,
\begin{equation}
-\frac{1}{4}\hmn\hmnup = -\frac{1}{4\ve^2} \left[ \dmu (\ve A_\nu) - \dnu (\ve A_\mu) \right]^2.
\end{equation}
Expanding and keeping terms to lowest order in $\ve$ and $\varphi$ give
\begin{align}
&-\frac{1}{4}\hmn\hmnup  \nonumber \\
&\simeq  -\frac{1}{4}\fmn\fmnup - \frac{1}{2\ve} \left[ (\dmu\ve) A_\nu - (\dnu\ve) A_\mu \right] \fmnup  \nonumber \\
&= -\frac{1}{4}\fmn\fmnup -\frac{1}{\ve}  (\dmu\ve) A_\nu  \fmnup  \nonumber \\
&\simeq  -\frac{1}{4}\fmn\fmnup -  \dmu\varphi \, A_\nu  \fmnup .
\end{align}
Because we start with a gauge invariant theory, the above Lagrangian is also gauge invariant. However, 
neither the kinetic term $-(1/4)\fmn\fmnup$ nor the interaction term $\dmu\varphi A_\nu  \fmnup$ is gauge invariant 
by themselves. Indeed, the field-strength tensor transforms under the gauge transformation (\ref{eq:gaugetransf}) as
$
\fmn \to \fmn - {\ve^{-2}} \left[ \dmu\ve \, \dnu\alpha - \dnu\ve \, \dmu\alpha\right],
$
which for constant $\ve=1$ reduces to the QED result. The corresponding transformations for the kinetic and interaction terms, ignoring all higher order terms in $\ve$, are given by
\begin{align}
-\frac{1}{4} \fmn\fmnup \to -\frac{1}{4} \fmn\fmnup + \frac{1}{\ve^2} (\dmu\ve)(\dnu\alpha)\fmnup
\end{align}
\begin{align}
-\frac{1}{\ve}  (\dmu\ve) A_\nu  \fmnup \to 
&-\frac{1}{\ve}  (\dmu\ve) A_\nu  \fmnup  \nn\\
&- \frac{1}{\ve^2} (\dmu\ve)(\dnu\alpha)\fmnup ,
\end{align}
and the sum of these terms, and therefore the Lagrangian, is gauge invariant, but not the individual terms. 
This is somewhat unusual, but shows that the interaction of the photon with the scalar is intimately connected to the kinetic term.

We now switch from the dimensionless field $\varphi$ to the canonically normalized field $\phi=\Lambda\varphi$.
The new scalar $\phi$ also couples to charged particles. The electric charge times photon field $\widetilde{e}A_\mu$ is to be everywhere replaced by $e\ve A_\mu$, so terms coupling $\phi$ to charged particles will be generated from the covariant derivatives for charged fermions.

The covariant derivative acting on charged fermions is given by $\widehat D_\mu = \dmu - i \widetilde{e}Q A_\mu$, where $Q$ is the charge of the corresponding fermion. Using $\widetilde{e}=e\ve$  gives 
\begin{equation}
\widehat D_\mu = \dmu - i e Q A_\mu - \frac{i e Q}{\Lambda} \phi A_\mu.
\end{equation}
If we now define $D_\mu = \dmu- i e Q A_\mu$ as the more familiar covariant derivative of electromagnetism,
the gauge invariant kinetic term for fermions is given by
\begin{equation}
{\cal L} \supset i\overline\psi \gamma^\mu \widehat D_\mu \psi
= i\overline\psi \gamma^\mu  D_\mu \psi + \frac{e Q}{\Lambda} \phi\, \overline\psi \gamma^\mu \psi \, A_\mu ,
\end{equation}
where the last term is a four-point interaction term for the vertex $\phi\gamma\bar\psi\psi$. This is the only direct coupling of the scalar field to fermions.

We finally allow the scalar $\phi$ to be massive. We then have the Lagrangian 
\begin{align}
{\cal L} &\supset \half(\dmu\phi)^2 - \half m^2\phi^2  \nn \\
&- \frac{1}{\Lambda} \dmu\phi \, A_\nu  \fmnup
+ \frac{e Q}{\Lambda} \phi\, \overline\psi \gamma^\mu \psi \, A_\mu,
\label{eq:Lag}
\end{align}
where  $\psi$ is a generic fermion field with charge $Q$.
Note that heuristically, the new interaction vertices of this model are obtained by inserting a scalar into each possible QED vertex and propagator.

\section{Alternative form of the model}
\label{app:alternative}

The interaction term derived above,
\begin{equation}
-\frac{1}{\ve}  (\dmu\ve) A_\nu  \fmnup
\simeq -\frac{1}{\Lambda} \dmu\phi \, A_\nu  \fmnup ,
\end{equation}
can be rewritten by performing an integration by parts,
\begin{equation}
-\frac{1}{\Lambda} \dmu\phi \, A_\nu  \fmnup 
\to  \frac{1}{\Lambda} \phi A_\nu \, \dmu \fmnup + \frac{1}{2\Lambda}\phi\fmn\fmnup ,
\end{equation}
ignoring total derivatives.
The first term on the rhs contains $\dmu \fmnup$, which by the equation of motion of the field (the Maxwell's equations) is given by $\dmu \fmnup=j^\nu= - e Q \overline\psi \gamma^\nu \psi $, and will therefore exactly cancel the interaction term $(e Q/\Lambda) \phi\, \overline\psi \gamma^\nu \psi \, A_\nu$. The second term on the rhs is the new form of the scalar-photon-photon vertex.

In the alternative form of the Lagrangian, we have a different expression for the scalar-photon coupling, and the four-point interaction of the scalar with two fermions and a photon has vanished. In the alternative form, fermions only interact with the scalar through intermediate photons. This is illustrated by the Feynman diagram for the $\phi\to\gm ff$ decay shown in Fig.~\ref{fig:phi2yff} in Appendix~\ref{app:3bd} below. In the alternative form of the model, this decay is mediated by an intermediate photon, as shown in the diagram. In the original form of the interaction, instead, this decay is directly given by the vertex that originates from the original four-point interaction.

Thus, the two forms of the theory certainly look very different, with different Feynman rules for corresponding vertices in the two theories and even different vertices, but in fact the two formulations are equivalent and will give the same $S$-matrix elements for any combination of initial and final states.

For example, it is straightforward to see that any process that only involves
real emission of photons through this interaction, and no virtual photon propagators, will have 
the same amplitude in both theories. This happens because when the condition of real, transverse photons 
is imposed, the Lorentz structures of the vertex in the two theories become identical. 

But the equivalence is guaranteed from more general arguments. Moving between the two descriptions is equivalent to a field redefinition, or a change of variables in the path integral. Such a field redefinition does not alter $S$-matrix elements, neither at tree-level or at higher orders. This is discussed in, e.g., Refs.~\cite{Chisholm:1961tha,Kamefuchi:1961sb,Kallosh:1972ap,Politzer:1980me,Arzt:1993gz}.
Indeed, this freedom is used to find the minimal set of operators in effective field theories; see, e.g.,~\cite{Brivio:2017vri} for a discussion.

\section{Three-body decay}
\label{app:3bd}

\begin{figure}[!h]
\includegraphics[height=3.5cm,width=4.5cm]{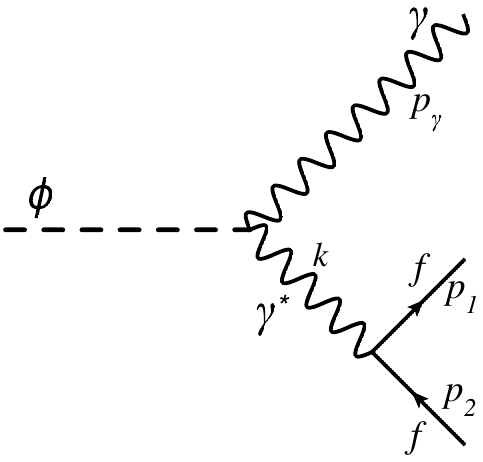}
\caption{Feynman diagram for the $\phi\to\gm ff$ decay.}
\label{fig:phi2yff}
\end{figure}

We present here the analytical expressions that are used to compute the three-body decay widths of $\phi_i$
as discussed in Sec.~\ref{ssec:BR}. 
For this, we consider the generic interactions given in Eq.~\eqref{eq:QEDlag2}. The three-body decay is mediated
by an off-shell photon as shown in Fig.~\ref{fig:phi2yff}. The four-momenta of the outgoing photon, the intermediate
photon, outgoing fermion, and outgoing antifermion are $p_\gm$, $k$, $p_1$, and $p_2$, respectively. To simplify the
computation, we choose a frame where the $ff$ pair is at rest and the invariant mass of the $ff$ pair is
$k^2=(p_1+p_2)^2=s$. The cosine of the angle between $\vec{p}_\gm$ and $\vec{p}_1$ is $c_\theta$.
The partial width of the $\phi\to\gm ff$ decay is given by
\begin{align}
\Gm_{\gm ff} &= \frac{N_c e^2Q_f^2}{512\pi^3 m_{\phi}^3}\int^{m_{\phi}^2}_{4m_f^2}ds\int_{-1}^1dc_\theta \nn\\
&\times\sqrt{1-\frac{4m_f^2}{s}}\lt(m_{\phi}^2 - s\rt)\lt|\mc{M}_{f}\rt|^2,
\end{align}
where $m_f$ is the mass of the fermion $f$ and its EM charge is $eQ_f$.
The amplitude squared of the process $\phi\to\gm ff$ is denoted by $|\mc{M}_f|^2$, which is given by
\begin{align}
|\mc{M}_{f}|^2 &=\frac{8e^2}{\Lm^2s^2}\big\{s(m_\phi^2 - s) (p_\gm\cdot p_1+p_\gm\cdot p_2)\nn\\
&- 4s(p_\gm\cdot p_1)(p_\gm\cdot p_2) + m_f^2(m_\phi^2 - s)^2\big\},
\end{align}

where $p_\gm\cdot p_1$ and $p_\gm\cdot p_2$ are given by
\begin{align}
p_\gm\cdot p_1 &= \frac{m_\phi^2 - s}{4}\lt(1 - \sqrt{1-\frac{4m_f^2}{s}}c_\theta\rt)\nn\\
p_\gm\cdot p_2 &= \frac{m_\phi^2 - s}{4}\lt(1 + \sqrt{1-\frac{4m_f^2}{s}}c_\theta\rt).
\end{align}

\bibliography{reference}{}
\bibliographystyle{JHEP}

\end{document}